\documentclass[aps,prresearch,twocolumn,showpacs,nofootinbib,floatfix,superscriptaddress,showpacs,longbibliography]{revtex4-2}
\usepackage[T1]{fontenc}
\usepackage{graphicx}% Include figure files
\usepackage{graphics}
\usepackage{dcolumn}% Align table columns on decimal point
\usepackage{bm}% bold math
\usepackage{epstopdf}
\usepackage{mathrsfs}
\usepackage{amssymb}
\usepackage{amsmath}
\usepackage[pdftex,colorlinks=true,linktocpage=true,linkcolor=blue,citecolor=blue]{hyperref}
\usepackage{xspace}
\usepackage{slashed}
\usepackage{nicefrac}
\usepackage{xcolor}
\usepackage[normalem]{ulem}
\usepackage{appendix}

\newcommand{\beq}{\begin{eqnarray}}
\newcommand{\eeq}{\end{eqnarray}}
\newcommand{\be}{\begin{eqnarray*}}
\newcommand{\ee}{\end{eqnarray*}}
\newcommand{\bqa}{\begin{eqnarray}}
\newcommand{\eqa}{\end{eqnarray}}

\begin{document}

%\title{Alternative order parameters for Potts model inspired from a machine learning study}
\title{Discovering quasiorder parameters in the Potts model:\\ A bridge between machine learning and critical phenomena}
% \title{Discovering alternative order parameters in the Potts model: a bridge between machine learning and critical phenomena}
\author{Yi-Lun Du}
\email{yilun.du@iat.cn}
\affiliation{Shandong Institute of Advanced Technology, Jinan 250100, China}
% \affiliation{Department of Physics, University of Oslo, Sem Sælands vei 24, Oslo 0371, Norway}
% \affiliation{Department of Physics and Technology, University of Bergen, Postboks 7803, 5020 Bergen, Norway}
%\affiliation{Frankfurt Institute for Advanced Studies, Giersch Science Center, D-60438 Frankfurt am Main, Germany}
%\affiliation{Institute f$\ddot{u}$r Theoretische Physik, Goethe Universit$\ddot{a}$t Frankfurt, D-60438 Frankfurt am Main, Germany}
%\affiliation{Department of Physics, Nanjing University, Nanjing 210093, China}

\author{Nan Su}
\email{nansu@fias.uni-frankfurt.de}
\affiliation{Frankfurt Institute for Advanced Studies, 60438 Frankfurt am Main, Germany}
%\affiliation{%
%J\"ulich Supercomputing Centre, Forschungszentrum J\"ulich, 52428 J\"ulich, Germany}

\author{Konrad Tywoniuk}
\email{konrad.tywoniuk@uib.no}
\affiliation{Department of Physics and Technology, University of Bergen, Postboks 7803, 5020 Bergen, Norway}

%\preprint{BI-TP 2014/12, ICCUB-14-058}

\date{\today}% It is always \today, today,

\begin{abstract}
Machine-learning (ML) models trained on Ising spin configurations have demonstrated surprising effectiveness in classifying phases of Potts models, even when processing severely reduced representations that retain only two spin states. To unravel this remarkable capability, we identify a family of alternative order parameters for the $q=3$ and $q=4$ Potts models on a square lattice, constructed from the occupancies of secondary and minimal spin states rather than the conventional dominant-state order parameter. Through systematic finite-size scaling analyses, we demonstrate that these quantities, along with a magnetization-like quantity derived from a reduced spin representation, accurately capture critical behavior, yielding critical temperatures and exponents consistent with established theoretical predictions and numerical benchmarks. Furthermore, we rigorously establish the fundamental relationships between these alternative (quasi)order parameters, demonstrating how they collectively encode criticality through different aspects of spin configurations. Our results clarify, within this specific setting, how reduced spin representations can retain the essential thermodynamic information needed for identifying critical behavior. Taken together, this work establishes a concrete bridge between Ising-trained ML models and critical phenomena in Potts systems by showing that Potts criticality can be encoded in more compact, non-traditional forms, thereby opening avenues for discovering analogous order parameters in broader spin systems.

% Our results provide a thermodynamic basis for ML's success in reduced representations. This work bridges data-driven ML approaches with fundamental statistical mechanics, showing that criticality in Potts systems can be encoded in more compact, non-traditional forms, and opens avenues for discovering analogous order parameters in broader spin systems.
\end{abstract}

% \pacs{12.38.Aw, 11.10.Wx, 12.38.Mh, 25.75.Nq}

\maketitle

\section{Introduction}
\label{sec: Intro}
Recent advances in machine learning (ML) have profoundly impacted the scientific studies~\cite{jordan2015machine,mehta2019high,carleo2019machine}. In condensed matter and statistical physics, ML has enabled tasks such as phase classification~\cite{wang2016discovering,van2017learning,wetzel2017unsupervised,hu2017discovering,carrasquilla2017machine,wetzel2017machine,rem2019identifying,dong2019machine,zhang2019machine,zhang2020interpreting,boyda2021finding}, topological feature identification~\cite{zhang2017quantum,zhang2018machine,araki2019phase,rodriguez2019identifying,zhang2019machine,zhang2020interpreting}, ground-state searching~\cite{carleo2017solving,mills2020finding,fan2023searching,djuric2025spin}, and accelerated Monte-Carlo sampling~\cite{huang2017accelerated,wang2017exploring,song2019accelerated,li2019accelerating,mcnaughton2020boosting}. Central to these applications is the ability of artificial neural networks (ANNs) to recognize patterns in high-dimensional data, such as spin configurations generated via Monte-Carlo simulations. Both supervised ~\cite{carrasquilla2017machine,wetzel2017machine,tian2023machine} and unsupervised ~\cite{wang2016discovering,van2017learning,wetzel2017unsupervised,hu2017discovering,ShibaFunai:2018aaw} approaches have proven successful in analyzing such simulation data. The Ising model has emerged as a key test bed for these ML techniques due to its simplicity and well-characterized critical behavior, with ANNs successfully determining its critical temperature and exponents~\cite{carrasquilla2017machine,van2017learning,kashiwa2019phase,li2019extracting,abuali2025deep}, establishing a paradigm for data-driven investigations of phase transitions. 

Despite these successes, the ``black-box'' nature of ANNs poses a significant challenge: Their decision-making processes and learned features often lack interpretability, limiting both reliability and generalizability. Recent work has begun addressing these limitations by examining the underlying mechanisms behind critical temperature determination~\cite{kim2018smallest} and exploring generalization in ML applications~\cite{westerhout2020generalization,CORTE2021110702}. These concerns become particularly salient when extending ML models to systems beyond their training data, such as from the Ising model to Potts model—a generalization of the former from Z(2) to Z($q$) symmetry. The Potts model serves as an ideal framework to explore the applicability~\cite{li2018applications,zhao2019machine,tan2020comprehensive,tirelli2022unsupervised,chen2023study,chen2024applications} and generalizability~\cite{shiina2020machine,bachtis2020mapping,fukushima2021can,giataganas2022neural,yau2022generalizability,tseng2024learning} of the trained networks across physical systems, as its $q$-component spin configurations differ fundamentally from the binary Ising spins. Furthermore, the universality classes of the Potts model also govern the phase transition of SU($N$) gauge theories, so this line of study also has direct impact to high-energy physics.

This fundamental difference necessitates careful reconciliation between Ising and Potts spin representations when studying generalizability. Recent work has proposed various procedures to bridge this gap~\cite{shiina2020machine,bachtis2020mapping,fukushima2021can,yau2022generalizability}. In Ref.~\cite{yau2022generalizability}, a mapping procedure of spin configuration from $q$ components to first two components, i.e., $\{1, 2, 3, \dots, q\}\mapsto\{-1, 1, 0, \dots, 0\}$, is employed without loss of generality. Remarkably, ANNs trained on Ising data can classify phases and extract critical exponents from these mapped Potts configurations for different $q$ value and lattice geometry~\cite{yau2022generalizability}, suggesting that critical properties persist in these simplified spin representations. Yet the fundamental reasons for this success, specifically what physical quantities the networks actually learn from reduced configurations, remain unresolved.

In this work, we address this gap by identifying alternative (quasi)order parameters for the Potts model that encode the model's critical behavior. Building on established evidence that ANNs trained on Ising model configurations effectively learn magnetization-like features~\cite{carrasquilla2017machine,kim2018smallest}, we hypothesize that when applied to reduced Potts model representations, these networks similarly extract quasiorder parameters analogous to magnetization, specifically, quantities proportional to the sum of all spin values. Furthermore, we propose a family of quantities derived not only from dominant spin state (conventional order parameter) but also from secondary and minimal occupancies, revealing a hidden encoding of critical properties. Through finite-size scaling (FSS) analyses of 
$q=3$ and $q=4$ Potts models on square lattices, we demonstrate that these alternative (quasi)order parameters—including a magnetization analog matching the form of $Z_5$ clock model order parameters~\cite{vanderzande1987bulk} and representing a static version of dynamic Potts order parameters~\cite{fernandes2006alternative}—accurately reproduce established critical temperatures and exponents. Our results illustrate how the mapping procedure preserves the relevant thermodynamic information needed for identifying criticality in the Potts model, thereby explaining why machine-learning models trained on the Ising model can succeed in this setting. This demonstrates that critical behavior can remain encoded in more compact, non-traditional forms of representation.

% Our results provide a thermodynamic foundation for ML’s empirical success, demonstrating that criticality in Potts systems can be robustly encoded in more compact, non-traditional forms.

This paper is organized as follows: Section.~\ref{sec: models} gives a brief introduction to the Ising and Potts models, detailing their conventional order parameters and Monte-Carlo simulation methodology. In Sec.~\ref{sec: Alternative order parameter}, we propose alternative (quasi)order parameters, rigorously validate their critical behavior through finite-size scaling, and elucidate their connection to ML-compatible mappings of spin configurations. Finally, in Sec.~\ref{sec: conclusion} we summarize our key findings and discuss their implications for both statistical physics and machine-learning applications. Complementary finite-size scaling analyses supporting our main results are presented in the Appendix. 

\section{The Ising and Potts models}            
\label{sec: models}
The Hamiltonian of the ferromagnetic nearest-neighbor Ising model with vanishing external magnetic field reads 
\beq
H_{\mathrm{Ising}}=-J\sum_{\langle i,j\rangle}\sigma_i\sigma_j, 
\eeq
where $\sigma\in\{-1, 1\}$ denotes the spin configurations, $\langle i,j\rangle$ stands for the nearest neighboring sites $i$ and $j$, and the coupling $J$ is set to unity. This system undergoes a second-order phase transition at the critical temperature $T_c=2/\ln(1+\sqrt{2})\approx 2.269$. The order parameter is the magnetization per site: $M=|\Sigma_i \sigma_i|/N$, where $N$ is the total number of spins. Below $T_c$, the system is ordered with $\langle M\rangle>0$, while above $T_c$, the system is disordered with $\langle M\rangle=0$.

The Potts model is a generalization of the Ising model from two to $q$ states. The Hamiltonian of ferromagnetic Potts model reads 
\beq
H_{\mathrm{Potts}}=-J\sum_{\langle i,j\rangle}\delta_{\mathrm{Kr}}(\sigma_i,\sigma_j),
\eeq
where $\sigma\in\{1,\dots,q\}$ and the Kronecker delta $\delta_{\mathrm{Kr}}$ evaluates to $1$ if $\sigma_i=\sigma_j$ and $0$ otherwise. For $q=2$, this reduces to the Ising model up to a factor of 2 in the coupling constant. The Ising model and Potts model can be defined for different lattice geometries and dimensions. For two-dimensional ($d=2$) $q$-state Potts model on a square lattice, the critical temperature $T_c=1/\ln{(1+\sqrt{q})}$~\cite{beffara2012self}, with second-order phase transitions for $q\leqslant4$~\cite{duminil2017continuity} and first order for $q\geqslant5$~\cite{duminil2016discontinuity}; see also the review~\cite{wu1982potts} and references therein. In this work, we restrict our attention to $q\leqslant4$ Potts model on a square lattice in two dimensions ($d=2$). The order parameter for $q$-state Potts model is usually defined as 
\beq
M_{\mathrm{Potts}} = \frac{q\frac{\mathrm{max}\{N_i\}}{N}-1}{q-1},
\label{eq:OP}
\eeq
where $N_i$ ($i=1, 2, \dots, q$) is the number of spins in state $i$ in the configuration of the Potts model and is referred to as the multiplicity of state $i$. $N=\sum_i^q N_i$ is the total number of spins and equals to $L^d$ with $L$ the lattice size. This definition actually makes use of the occupancy of the most numerous state in each configuration $\mathrm{max}\{N_i\}/N$. For $q=2$, it is equivalent to the conventional order parameter of the Ising model, i.e., the global magnetization
% \kt{Is this the analogy of the global magnetization?}

\beq
M_{\mathrm{Ising}} = \frac{|N_1-N_2|}{N},
\label{eq:OPIsing}
\eeq
where $N_1$ and $N_2$ are referred to as the number of spins $-1$ and $+1$, respectively. $N_1/N$ and $N_2/N$ are not independent and we actually have

\beq
\frac{\mathrm{max}\{N_i\}}{N} = \frac{|N_1-N_2|}{2N}+\frac{1}{2}.
\label{eq: Ising-Potts}
\eeq
Since the definition of $M_{\mathrm{Potts}}$ in Eq.~(\ref{eq:OP}) contains the maximal occupancy, in the following we use $M_{\mathrm{max}}$ to label $M_{\mathrm{Potts}}$ to distinguish the alternative (quasi)order parameters we will propose for the Potts model. 

% \section{Generation of spin configurations}

In our Monte Carlo simulations, we use the Swendsen-Wang algorithm~\cite{swendsen1987nonuniversal} by the package in GitHub~\cite{Swendsen-Wang_Potts_algorithm} to generate spin configurations with periodic boundary conditions. Each lattice is first equilibrated for $1000$ cluster updates, after which spin configurations are sampled every $k$ cluster updates, where $k$ is the number of updates to decorrelate the current update with previous ones for each lattice size setup. A total of 64 sampling temperatures $T$ are used in Monte Carlo simulations with an interval $0.002J$, and the critical temperature $T_c$ is covered. For lattice of size $L\times L$ with $L=\{16, 32, 48, 64, 80, 96\}$, 100\,000 samples are generated for each temperature.

Taking $q=3$ Potts model on a square lattice ($d=2$) as an example, we show the simulation results of the order parameter, i.e., the ensemble average of $M_{\mathrm{max}}(T)$ for different lattice size $L$ in Fig.~\ref{Fig: Mmax} in the Appendix. Theoretically, $\bar{M}_{\mathrm{max}}(T)$ satisfies that $\bar{M}_{\mathrm{max}}=0$ when $T>T_c$ and $\bar{M}_{\mathrm{max}}>0$ when $T<T_c$. The Monte-Carlo lattice simulation shows the finite-size effect for this picture in the left panel. Near $T_c$, the order parameter $\bar{M}$ in general satisfies $\bar{M}(L, T)=L^{-\Delta_\sigma}\tilde{m}(tL^{1/\nu})$, where $t=(T-T_c)/J$, $\nu=5/6$, and $\Delta_\sigma=2/15$ are the critical exponent and scaling dimension~\cite{wu1982potts}. With the finite-size scaling procedure, the plot $\bar{M}(L, T)L^{\Delta_\sigma}-T$ with different lattice size $L$ shows a crossing point at $T_c$ and the plot $\bar{M}(L, T)L^{\Delta_\sigma}-tL^{1/\nu}$ with different $L$ presents a data collapse on top of each other, as shown in the middle and right panels of Fig.~\ref{Fig: Mmax} in the Appendix, respectively. 

\section{Alternative (quasi)order parameters for Potts model}
\label{sec: Alternative order parameter}

\begin{figure}[!htb]
\centering
\includegraphics[width=3.1in]{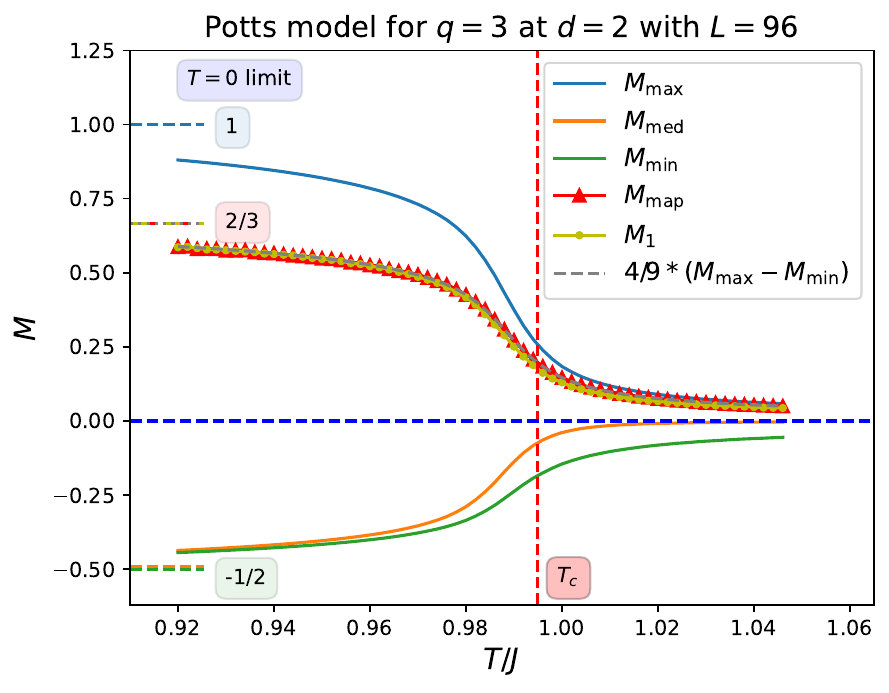}
\includegraphics[width=3.1in]{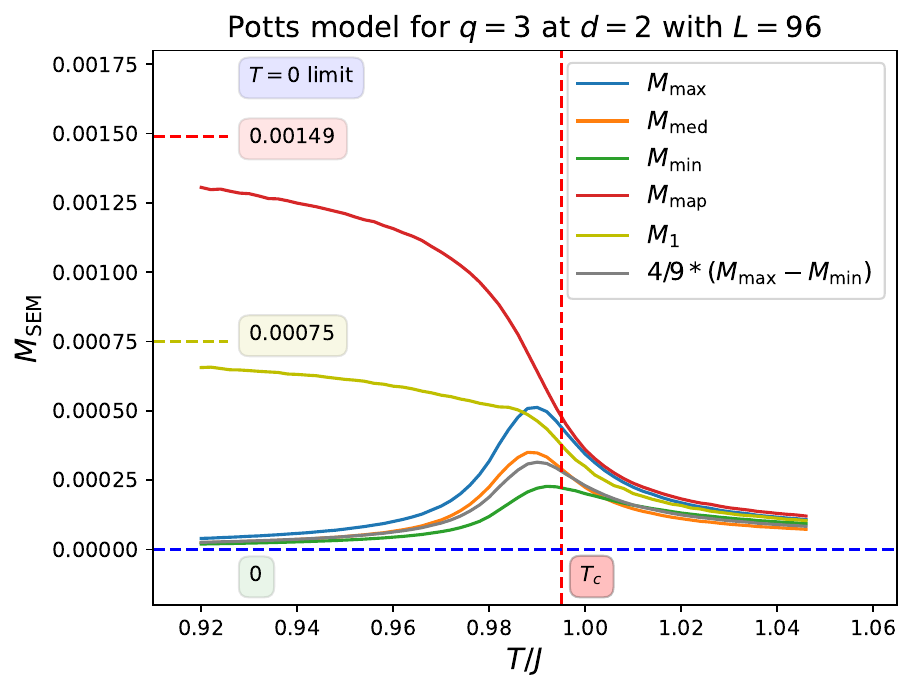}
\includegraphics[width=3.1in]{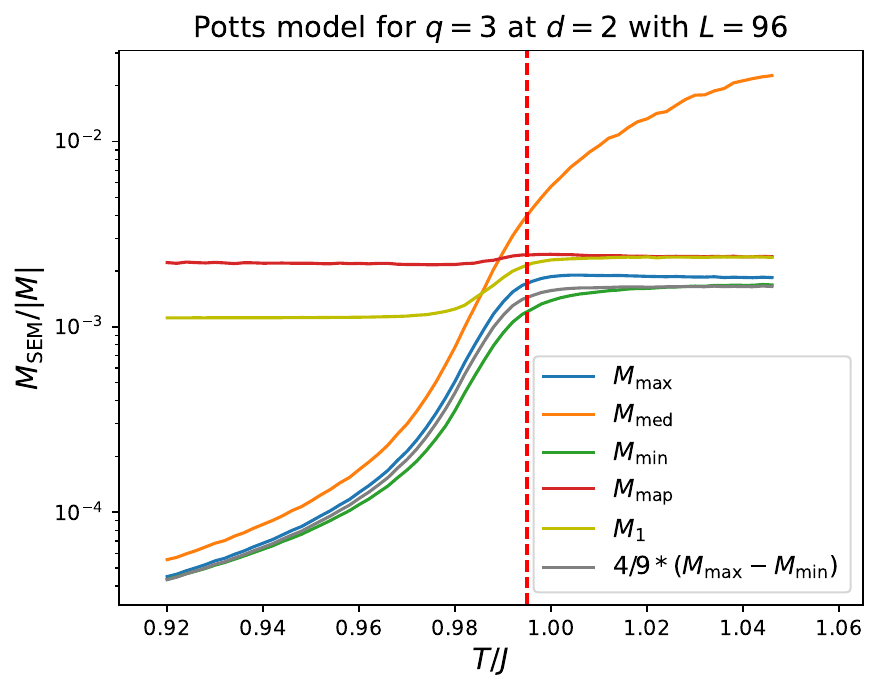}
\caption{(Color online) The mean (upper), standard error of mean (SEM) (middle), and relative standard error (RSE) (lower) of alternative order parameters $M_{\mathrm{max}}$, $M_{\mathrm{med}}$, and $M_{\mathrm{min}}$ and quasiorder parameter $M_{\mathrm{map}}$ as functions of temperature $T$ for the three-state Potts model at $d=2$, simulated on a square lattice of size $L=96$. The relation $M_{\mathrm{map}}=\frac{4}{9}(M_{\mathrm{max}}-M_{\mathrm{min}})$ is confirmed by the overlap of the curves in the upper panel. The blue horizontal dashed lines mark zero to indicate the sign, while the red vertical dashed lines denote the theoretical critical temperature $T_c$.}
\label{Fig: all op}
\end{figure}

Reference~\cite{yau2022generalizability} demonstrated a remarkable capability of ANNs in classifying phases across different spin models. ANNs were first trained using Ising model spin configurations across various temperatures to classify phases. Then, these Ising-trained ANNs were successfully applied to classify phases in Potts models of the same lattice size. To bridge the fundamental difference in spin representations between these models (binary spins in Ising versus $q$-state spins in Potts), an elegant mapping procedure $\{1, 2, 3, \dots, q\}\mapsto\{-1, 1, 0, \dots, 0\}$ was introduced for $q \leqslant 7$, where only first two spin states are preserved while others are set to zero. This mapping maintained all essential physics while providing compatibility with the Ising-trained networks. Remarkably, these ANNs not only accurately classified Potts model phases but also correctly extracted critical exponents for $q \leqslant 4$ cases. The success of this mapping procedure inspired our current investigation into alternative order parameters that might underlie this surprising generalizability.

\subsection{$q=3$ Potts model on a square lattice}

\begin{table*}[tbh]
\small
\centering
% \begin{tabular}{|c|c|c|c|}
\begin{tabular}{| >{\centering}p{3.5cm} | >{\centering}p{3.5cm} | >{\centering}p{2.8cm} |>{\centering\arraybackslash}p{2.8cm} |} % >{\centering\arraybackslash}p{2.8cm}|}
\hline
(Quasi)order parameter &  $T_c$ & $\Delta_\sigma$ & $\nu$ \\ % & 1/$\nu$\\
\hline
$M_{\mathrm{max}}$ &  0.9935(1) & 0.137(6)  & 0.84(2) \\ %&  1.19(2) \\
\hline
$M_{\mathrm{min}}$  & 0.9931(1) & 0.139(4)  & 0.83(6) \\ %& 1.21(8)\\
\hline
$M_{\mathrm{med}}$  & 0.9940(2) & 0.138(16) & 0.84(5) \\% & 1.19(7)\\
\hline
$M_{\mathrm{map}}$  & 0.9942(4) & 0.134(16) & 0.84(10) \\ % & 1.19(13) \\
\hline
Theoretical value  & 1/$\ln{(1+\sqrt{3})}\sim 0.9950$ & $2/15 \sim 0.1333$  & $5/6 \sim 0.8333$  \\ % & 6/5=1.2\\
\hline
\end{tabular}
\caption{Critical properties of the three-state Potts model at $d=2$ on a square lattice, obtained using the alternative order parameters $M_{\mathrm{max}}$, $M_{\mathrm{med}}$, and $M_{\mathrm{min}}$ and quasiorder parameter $M_{\mathrm{map}}$. The critical temperature $T_c$, scaling dimension $\Delta_\sigma$, and critical exponent $\nu$, along with their standard errors, are estimated via finite-size scaling within the interval $T\in(0.920, 1.046)$ with step size $\Delta T=0.002$. Theoretical values from the literature are included for comparison~\cite{wu1982potts}.}  
\label{Table: q=3, shell fine}
\end{table*}

For the $q=3$ case, the mapping of three spin components takes the form
\beq
\{1, 2, 3\}\mapsto\{-1, 1, 0\},
\label{eq: mapping}
\eeq
where we keep the first two components and ignore the third one by setting it as zero. Building on the established understanding that ANNs trained on Ising models learn to detect magnetization patterns, i.e., the sum of all spin values, we define a corresponding quantity for the mapped Potts model:
\beq
M_{\mathrm{map}} = \frac{|N_1-N_2|}{N}.
\label{eq:OPmap}
\eeq
$M_{\mathrm{map}}$ represents a natural generalization of order parameter for the Ising model. For $q=2$ Potts model, $M_{\mathrm{map}}$ is equivalent to $M_{\mathrm{max}}$ in Eq.~(\ref{eq:OP}) as demonstrated by Eq.~(\ref{eq: Ising-Potts}). Our finite-size scaling analysis (shown in Fig.~\ref{Fig: Mmap} in the Appendix) demonstrates that $M_{\mathrm{map}}$ successfully captures critical behavior using the known critical exponents $\nu$ and $\Delta_\sigma$,  confirming its validity as an order parameter. In this work, we are concerned with the static critical properties of Potts model in equilibrium and notably $M_{\mathrm{map}}$ could be viewed as a static version of the alternative order parameter for Potts model at an early stage of the time evolution of dynamics at non-equilibrium~\cite{fernandes2006alternative}. It is also worth noting that $M_{\mathrm{map}}$ has the same form of the order parameter for $Z_5$ model, \textit{aka} five-state clock model~\cite{vanderzande1987bulk}.

To relate $M_{\mathrm{max}}$ and $M_{\mathrm{map}}$ theoretically, we can make use of the minimal state occupancy in each configuration $\mathrm{min}\{N_i\}/N$. As we know, $\mathrm{min}\{N_i\}/N$ also tends to $1/q$ when $T>T_c$. While at low $T$, this occupancy is suppressed by the dominant state. So, here we construct a quantity $M_{\mathrm{min}}$ by replacing the occupancy of the most state in each configuration $\mathrm{max}\{N_i\}/N$ in Eq.~(\ref{eq:OP}) with the least one, $\mathrm{min}\{N_i\}/N$:
\beq
M_{\mathrm{min}} = \frac{q\frac{\mathrm{min}\{N_i\}}{N}-1}{q-1}.
\label{eq:OPmin}
\eeq
We assume that in one configuration, the occupancy of three states descends as $a, b, c$ with $a+b+c=1$. With the Z(3) symmetry, we could have $A_3^3=6$ cases of how these three occupancies are assigned to these three states. After smearing a specific state of these three ones with the mapping procedure, the average $M_{\mathrm{map}}$ of these six configurations equals $2(a-c)/3$. Since $M_{\mathrm{max}}=(3a-1)/2$ and $M_{\mathrm{min}}=(3c-1)/2$, as in Eqs.~(\ref{eq:OP}) and (\ref{eq:OPmin}), the equilibrium averages satisfy the fundamental relationship 
\beq
\bar{M}_{\mathrm{map}} = \frac{4}{9}(\bar{M}_{\mathrm{max}}-\bar{M}_{\mathrm{min}}),
\label{eq:OPq3}
\eeq
where the overline signifies the ensemble average. This is also confirmed numerically in the upper panel of Fig.~\ref{Fig: all op}, where $M_{\mathrm{max}}$, $M_{\mathrm{min}}$, and $M_{\mathrm{map}}$ as functions of temperature $T$ with lattice size $L=96$ are shown for visual comparison and the curve of $M_{\mathrm{map}}$ labeled with red triangles and that of $\frac{4}{9}(M_{\mathrm{max}}-M_{\mathrm{min}})$ labeled with gray dashed line collapse on top of each other. In the middle and lower panels of Fig.~\ref{Fig: all op}, we show the standard error of mean (SEM) and relative standard error (RSE), i.e., the ratio of SEM over mean, of these quantities, respectively. Due to the high statistics in this work (100\,000 samples), the standard errors of mean are too small to be visualized along with their mean values in the upper panel. One can observe that below $T_c$ the fluctuations of $M_{\mathrm{map}}$ are much larger than those of other quantities, particularly $\frac{4}{9}(M_{\mathrm{max}}-M_{\mathrm{min}})$, which are mainly due to the non-thermal fluctuations between those aforementioned six symmetric configurations for $M_{\mathrm{map}}$ arising from the different treatment on all Potts states. 

One can easily understand these non-thermal fluctuations in $M_{\mathrm{map}}$ at $T=0$ limit, where one expects full occupation of one of the Potts states and zero occupation of the other states, i.e., $a=1$ and $b=c=0$. For $q = 3$, there are $A_3^3/A_2^2=3$ equally likely possibilities,
\beq
(N_1, N_2, N_3)/N = (1, 0, 0), (0, 1, 0), (0, 0, 1).
\eeq
For these three configurations, $M_{\mathrm{map}}$ takes the values 1, 1, and 0, despite that all three configurations are ordered to the same extent. This leads to a mean $\bar{M}_{\mathrm{map}} = 2/3$, a standard deviation (SD) $M_{\mathrm{map}}^{\mathrm{SD}}$ = $\sqrt{2}/3$, and a standard error of the mean $M_{\mathrm{map}}^{\mathrm{SEM}}$=$M_{\mathrm{map}}^{\mathrm{SD}}/\sqrt{100\,000}=0.00149$. Note that these predictions for $T = 0$ are quite compatible with the simulation results in the upper two panels of Fig.~\ref{Fig: all op}. For general $q$ states, $\bar{M}_{\mathrm{map}} = 2/q$, $M_{\mathrm{map}}^{\mathrm{SD}}$ = $\frac{\sqrt{2(q-2)}}{q}$, and $M_{\mathrm{map}}^{\mathrm{SEM}}$=$M_{\mathrm{map}}^{\mathrm{SD}}/\sqrt{n}$, where $n$ is the total sample number. 

%We also label the limits of both the mean values and the SEM for various order parameters in Fig.~\ref{Fig: all op}.

Note that the fluctuations of $M_{\mathrm{map}}$, quantified by its SEM, do not behave like a susceptibility with a clear peak near $T_c$ for finite system size. In this sense, $M_{\mathrm{map}}$ can only be regarded as a quasiorder parameter. However, as $T$ approaches $T_c$ from above in the second panel of Fig.~\ref{Fig: all op}, the SEM for $M_{\mathrm{map}}$ rises steeply, as if approaching a peak, similar to the behavior of $M_{\mathrm{max}}$ and $M_{\mathrm{min}}$. We interpret this as a susceptibility-like enhancement that, however, is difficult to resolve for $T<T_c$ because of additional non-thermal fluctuations inherent to the definition of $M_{\mathrm{map}}$, which increase with decreasing temperature, as the system orders increasingly, and become strongest at $T = 0$.

An even simpler quantity than $M_\mathrm{map}$ is
\begin{equation}
M_1=\frac{|q\frac{N_1}{N}-1|}{q-1}.
\end{equation}
It reduces to the Ising order parameter when $q=2$ due to $|\frac{2N_1}{N}-1|=\frac{|N_1-N_2|}{N}$ and is expected to behave as an order parameter in general for $q$-state Potts model. However, it does not treat all Potts states equally and therefore introduces non-thermal fluctuations as $M_\mathrm{map}$. This quantity is of interest since it only involves one of the Potts states and provides motivation for ANN studies to extract Potts critical properties from simplified configurations that only retain the occupancy of one of the $q$ states. We include the corresponding results for $M_1$ in Fig.~\ref{Fig: all op}. Interestingly, the mean values of $M_1$ coincide with $M_\mathrm{map}$ and its $T=0$ limit can be obtained analytically as $2/q$, consistent with that of $M_\mathrm{map}$ as well. The SEM of $M_1$, with the limit $\frac{q-2}{q\sqrt{q-1}}\frac{1}{\sqrt{n}}=0.000\,75$ at $T = 0$, is smaller than that of $M_\mathrm{map}$. More generally, we find that any quantity, such as $M_\mathrm{map}$ and $M_1$, which treats ordering in a subset of Potts states differently from the others, will have non-thermal fluctuations that grow as $T$ decreases and may obscure the susceptibility peak associated with the thermal fluctuations near the critical point.

With Eq.~(\ref{eq:OPq3}), we can conjecture that $M_{\mathrm{min}}$ could also serve as an alternative order parameter. For $q = 3$, we also have the medium state for each configuration and therefore define
\beq
M_{\mathrm{med}} = \frac{q\frac{\mathrm{medium}\{N_i\}}{N}-1}{q-1}.
\label{eq:OPmed}
\eeq 
Due to the constraint that $M_{\mathrm{max}}+M_{\mathrm{med}}+M_{\mathrm{min}}=1$, only two of them are independent. We therefore conjecture that $M_{\mathrm{med}}$ could serve as an alternative order parameter as well. The temperature dependence of $|M_{\mathrm{min}}|$ and $|M_{\mathrm{med}}|$ and their associated finite-size scaled results are shown, respectively, in Figs.~\ref{Fig: Mmin} and~\ref{Fig: Mmed} in the Appendix with the known values of $\nu$ and $\Delta_\sigma$, which validates that $|M_{\mathrm{min}}|$ and $|M_{\mathrm{med}}|$ could also serve as alternative order parameters. There we take the absolute value of $M_{\mathrm{min}}$ and $M_{\mathrm{med}}$ so that they increase monotonically as temperature decreases and reach a maximum at the ground state with $T=0$, similar to the traditional order parameter $M_{\mathrm{max}}$. Any order parameter that is symmetric in the occupancies of the $q$ states, such as $M_{\mathrm{max}}$, $M_{\mathrm{min}}$, or $M_{\mathrm{med}}$, takes the same value for each of the three ground-state configurations, and therefore exhibits no fluctuations at $T = 0$. Specifically, at this limit one finds $M_{\mathrm{max}} = 1$, $M_{\mathrm{min}} = M_{\mathrm{med}} = -1/2$, with both the SD and SEM vanishing. These predictions labeled by the short dashed lines are also quite compatible with the simulation results in the upper two panels of Fig.~\ref{Fig: all op}. 
From the lower panel of Fig.~\ref{Fig: all op}, it is obvious that although $M_{\mathrm{max}}$, $M_{\mathrm{med}}$, and $M_{\mathrm{min}}$ all could serve as the order parameters of Potts model, which is consistent with the model's universality class and symmetry, they are not completely equivalent in the sense of their relative standard error, and $M_{\mathrm{min}}$ exhibits the minimal relative standard error for the whole critical temperature interval of interest.

With the FSS procedure~\cite{melchert2009autoscale}, we locate the critical temperature and extract the critical exponents for all these order parameters $M_{\mathrm{max}}$, $M_{\mathrm{med}}$, and $M_{\mathrm{min}}$ and quasiorder parameter $M_{\mathrm{map}}$ as detailed in Table~\ref{Table: q=3, shell fine}. They agree with the theoretical values very well and we conclude that all occupancies of these states along with the magnetization-like quantity $M_{\mathrm{map}}$ can serve as (quasi)order parameters for $q=3$ Potts model. In terms of quantifying order, the suppression of secondary or minimally occupied spin states is directly complementary to the dominance of the majority spin type. This suppression reflects the onset of symmetry breaking and forms the conceptual basis for proposing these quantities as alternative order parameters. While these quantities may be mathematically related to  $M_{\mathrm{max}}$, they represent a distinct and meaningful perspective, i.e., focusing on the elimination of minority components rather than the rise of the dominant one. This conceptual shift provides complementary insight into the nature of order in the Potts models and helps to elucidate the features learned by machine-learning models that implicitly capture phase transitions.

It would be interesting to check whether the above observations and conclusion are valid for other $q$-state Potts models. In Ising model, we have two occupancies and only one is independent, which is a trivial case. So, we will turn to $q=4$ Potts model in next subsection. %\Du{Compare the accuracy from $M_{\mathrm{map}}$ and $M_{\mathrm{max}}$}

% \clearpage

\subsection{$q=4$ Potts model on a square lattice}

\begin{figure}[!htb]
\centering
\includegraphics[width=3.in]{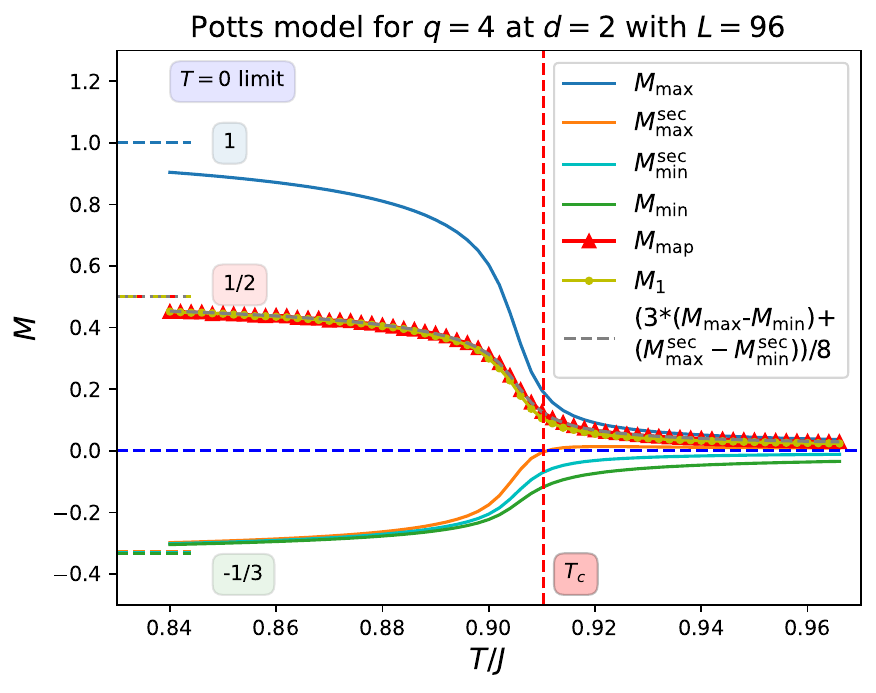}
\includegraphics[width=3.1in]{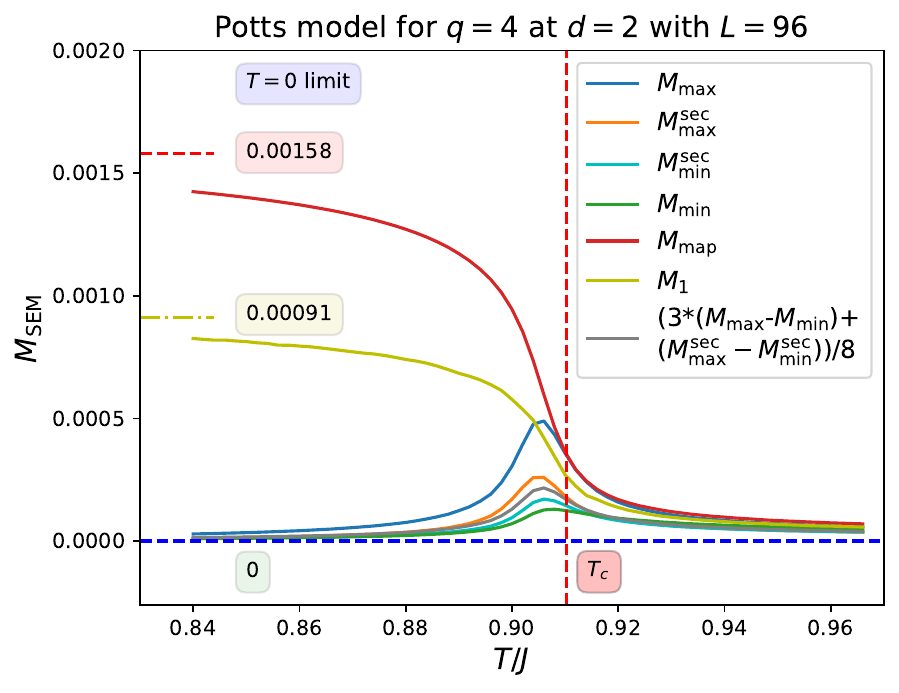}
\includegraphics[width=3.1in]{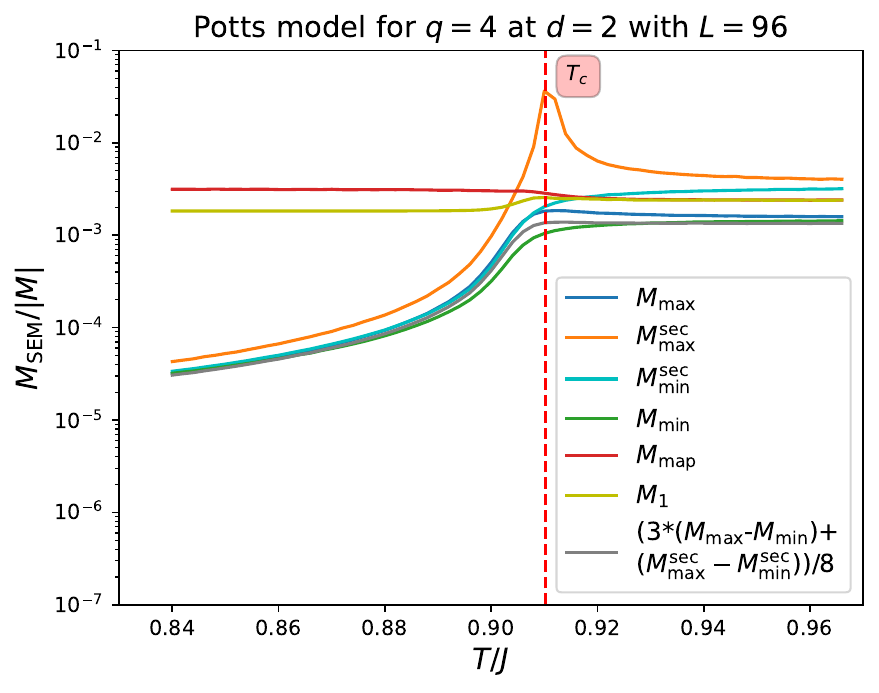}
\caption{(Color online) The mean (upper), standard error of mean (SEM) (middle), and relative standard error (RSE) (lower) of the alternative order parameters $M_{\mathrm{max}}^{}$, $M_{\mathrm{max}}^{\mathrm{sec}}$, $M_{\mathrm{min}}^{\mathrm{sec}}$, $M_{\mathrm{min}}^{}$, and quasiorder parameter $M_{\mathrm{map}}^{}$ as functions of temperature $T$ for the four-state Potts model at $d=2$, simulated on a square lattice of size $L=96$. The relation $M_{\mathrm{map}}=\frac{1}{8}(3(M_{\mathrm{max}}-M_{\mathrm{min}})+M_{\mathrm{max}}^{\mathrm{sec}}-M_{\mathrm{min}}^{\mathrm{sec}})$ is confirmed by the overlap of the curves in the upper panel. The blue horizontal dashed lines mark zero to indicate the sign, while the red vertical dashed lines denote the theoretical critical temperature $T_c$.}
\label{Fig: all op q4}
\end{figure}

For the $q = 4$ case, besides the most and least states, we also have the second most and second least states for each configuration; therefore, we define
\beq
M_{\mathrm{max}}^{\mathrm{sec}} = \frac{q\frac{\mathrm{secmax}\{N_i\}}{N}-1}{q-1},
\label{eq:OPsecmaxq4}
\eeq
\beq
M_{\mathrm{min}}^{\mathrm{sec}} = \frac{q\frac{\mathrm{secmin}\{N_i\}}{N}-1}{q-1}.
\label{eq:OPsecminq4}
\eeq
In the mapping procedure for $q=4$ case~\cite{yau2022generalizability}, only first two states are kept in the configuration without loss of generality, 
\beq
\{1, 2, 3, 4\}\mapsto\{-1, 1, 0, 0\},
\label{eq: mapping for q=4}
\eeq
and $M_{\mathrm{map}}$ is defined the same as in Eq.~(\ref{eq:OPmap}). As the analysis for three-state Potts model, we assume that in one configuration, the occupancy of four states descends as $a, b, c, d$ with $a+b+c+d=1$. With the Z(4) symmetry, we could have $A_4^4=24$ cases of how these four occupancies are assigned to these four states. After smearing two specific states of these four ones with the mapping procedure, the average $M_{\mathrm{map}}$ of these 24 configurations equals $(3(a-d)+(b-c))/6$. Since $M_{\mathrm{max}}=(4a-1)/3$, $M_{\mathrm{min}}=(4d-1)/3$, $M_{\mathrm{max}}^{\mathrm{sec}}=(4b-1)/3$, and $M_{\mathrm{min}}^{\mathrm{sec}}=(4c-1)/3$, as in Eqs.~(\ref{eq:OP}), (\ref{eq:OPmin}), (\ref{eq:OPsecmaxq4}), and (\ref{eq:OPsecminq4}), the equilibrium averages satisfy the fundamental relationship
\beq
\bar{M}_{\mathrm{map}} = \frac{1}{8}(3(\bar{M}_{\mathrm{max}}-\bar{M}_{\mathrm{min}})+\bar{M}_{\mathrm{max}}^{\mathrm{sec}}-\bar{M}_{\mathrm{min}}^{\mathrm{sec}}),
\label{eq:OPq4}
\eeq
where the overline signifies the ensemble average. This is also confirmed numerically in the upper panel of Fig.~\ref{Fig: all op q4}, where $M_{\mathrm{max}}^{}$, $M_{\mathrm{max}}^{\mathrm{sec}}$, $M_{\mathrm{min}}^{\mathrm{sec}}$, $M_{\mathrm{min}}^{}$, $M_{\mathrm{map}}^{}$, and $M_1^{}$ as functions of temperature $T$ with lattice size $L=96$ are shown for visual comparison and the curves of $M_{\mathrm{map}}$ labeled with red triangles, $M_1^{}$ labeled with yellow dots, and that of $\frac{1}{8}(3(M_{\mathrm{max}}-M_{\mathrm{min}})+M_{\mathrm{max}}^{\mathrm{sec}}-M_{\mathrm{min}}^{\mathrm{sec}})$ labeled with gray dashed line collapse on top of each other. In the middle and lower panels of Fig.~\ref{Fig: all op q4}, we show the SEM and RSE of these quantities, respectively. Similar to $q=3$ case, one can observe that below $T_c$ the fluctuations of $M_{\mathrm{map}}$ and $M_1$ are much larger than those of other quantities, particularly $\frac{1}{8}(3(M_{\mathrm{max}}-M_{\mathrm{min}})+M_{\mathrm{max}}^{\mathrm{sec}}-M_{\mathrm{min}}^{\mathrm{sec}})$, which are mainly due to the non-thermal fluctuations between those aforementioned 24 symmetric configurations for $M_{\mathrm{map}}$ arising from the different treatment on all Potts states. 

We estimate the non-thermal fluctuations in $M_{\mathrm{map}}$ and $M_1$ at $T=0$ limit where $a=1$ and $b=c=d=0$. For $q = 4$, there are $A_4^4/A_3^3=4$ equally likely possibilities,
\begin{gather}
(N_1, N_2, N_3, N_4)/N \nonumber \\
=(1, 0, 0, 0), (0, 1, 0, 0), (0, 0, 1, 0), (0, 0, 0, 1)
\end{gather}
For these four configurations, $M_{\mathrm{map}}$ takes the values 1, 1, 0, and 0, despite that all four configurations are ordered to the same extent. This yields $\bar{M}_{\mathrm{map}} = 1/2$, $M_{\mathrm{map}}^{\mathrm{SD}}$ = $1/2$, and $M_{\mathrm{map}}^{\mathrm{SEM}}$=$M_{\mathrm{map}}^{\mathrm{SD}}/\sqrt{100\,000}=0.001\,58$. While $M_1$ takes the values 1, 1/3, 1/3, and 1/3, leading to $\bar{M}_1 = 1/2$, $M_1^{\mathrm{SD}}$ = $\sqrt{1/12}$, and $M_1^{\mathrm{SEM}}$=$M_1^{\mathrm{SD}}/\sqrt{100\,000}=0.000\,91$. These $T = 0$ predictions are fully consistent with the general
$q$-state results discussed in the previous subsection and are quite compatible with the simulation data shown in the upper two panels of Fig.~\ref{Fig: all op q4}. Again, the fluctuations of $M_{\mathrm{map}}$ and $M_1$, quantified by their SEM, do not behave like a susceptibility with a peak near the $T_c$ for finite system size. In this sense, $M_{\mathrm{map}}$ and $M_1$ can only be regarded as quasiorder parameters. 

Order parameters that are symmetric in the occupancies of the $q$ states, such as $M_{\mathrm{max}}$, $M_{\mathrm{min}}$, $M_{\mathrm{max}}^{\mathrm{sec}}$, or $M_{\mathrm{min}}^{\mathrm{sec}}$, take the same value for each of the four ground-state configurations, and therefore exhibit no fluctuations at $T = 0$. Specifically, at this limit one finds $M_{\mathrm{max}} = 1$, $M_{\mathrm{min}} = M_{\mathrm{max}}^{\mathrm{sec}}= M_{\mathrm{min}}^{\mathrm{sec}} = -1/3$, with both the SD and SEM vanishing. These predictions labeled by the short dashed lines are also quite compatible with the simulation results in the upper two panels of Fig.~\ref{Fig: all op q4}.

\begin{table*}[tbh]
\small
\centering
% \begin{tabular}{|c|c|c|c|}
\begin{tabular}{| >{\centering}p{3.5cm} | >{\centering}p{3.5cm} | >{\centering}p{2.8cm} |>{\centering\arraybackslash}p{2.8cm} |} % >{\centering\arraybackslash}p{2.8cm}|}
\hline
(Quasi)order parameter &  $T_c$ & $\Delta_\sigma$  & $\nu$ \\ % & $1/\nu$\\
\hline
$M_{\mathrm{max}}^{}$ & 0.909\,48(8)  & 0.128(12)   & 0.723(27) \\ % & 1.384(50) \\
\hline
$M_{\mathrm{max}}^{\mathrm{sec}}$ & 0.909\,65(8) & 0.128(42) &  0.724(5)  \\ %  & 1.382(9)\\
\hline
$M_{\mathrm{min}}^{\mathrm{sec}}$  & 0.909\,45(8)  & 0.128(15)  & 0.721(3) \\ % & 1.386(5) \\
\hline
$M_{\mathrm{min}}^{}$  &  0.909\,16(8)  & 0.129(8)  & 0.728\,12(1) \\ % & 1.37339(1) \\
\hline
$M_{\mathrm{map}}^{}$   & 0.9094(2)  & 0.128(22) & 0.722(13) \\ % & 1.385(24)   \\
\hline
\, Numerical estimate*  & 1/$\ln{(1+\sqrt{4})}\sim 0.9102$ & 0.128* &  0.722* \\% & 1.385* \\
\hline
Theoretical value  & 1/$\ln{(1+\sqrt{4})}\sim 0.9102$ & $1/8=0.125$ &  $2/3~\sim 0.667$ \\
\hline
\end{tabular}
\caption{Critical properties of the four-state Potts model at $d=2$ on a square lattice, obtained using the alternative order parameters $M_{\mathrm{max}}^{}$, $M_{\mathrm{max}}^{\mathrm{sec}}$, $M_{\mathrm{min}}^{\mathrm{sec}}$, and $M_{\mathrm{min}}^{}$ and quasiorder parameter $M_{\mathrm{map}}^{}$. The critical temperature $T_c$, scaling dimension $\Delta_\sigma$, and critical exponent $\nu$, along with their standard errors, are estimated via finite-size scaling within the interval $T\in(0.840, 0.966)$ with step size $\Delta T=0.002$. For comparison, we include Monte-Carlo numerical estimates of $\Delta_\sigma$ and $\nu$ that neglect higher-order corrections and mark them with *~\cite{salas1997logarithmic}, along with the theoretical values from the literature~\cite{wu1982potts}.}  
\label{Table: q=4, shell fine}
\end{table*}

The temperature dependence of $M_{\mathrm{max}}^{}$, $-M_{\mathrm{max}}^{\mathrm{sec}}$, $|M_{\mathrm{min}}^{\mathrm{sec}}|$, $|M_{\mathrm{min}}^{}|$, and $M_{\mathrm{map}}^{}$ and their associated finite-size scaled results are shown, respectively, in Figs.~\ref{Fig: Mmaxq4}$-$\ref{Fig: Mmapq4} in the Appendix with the known Monte-Carlo numerical estimates of $\nu=0.722$ and $\Delta_\sigma=0.128$~\cite{salas1997logarithmic}, which validates that all of these quantities could serve as alternative (quasi)order parameters. There we take the absolute value or reverse the sign of $M_{\mathrm{min}}$, $M^{\mathrm{sec}}_{\mathrm{min}}$ and $M^{\mathrm{sec}}_{\mathrm{max}}$ so that they increase monotonically as temperature decreases and reach a maximum at the ground state with $T=0$, similar to the traditional order parameter $M_{\mathrm{max}}$. From the lower panel of Fig.~\ref{Fig: all op q4}, it is obvious that although $M_{\mathrm{max}}$, $M_{\mathrm{max}}^{\mathrm{sec}}$, $M_{\mathrm{min}}^{\mathrm{sec}}$, and $M_{\mathrm{min}}$ all could serve as the order parameters of Potts model, they are not completely equivalent in the sense of their relative standard error, and same as $q=3$ case, $M_{\mathrm{min}}$ exhibits the minimal relative standard error for the whole critical temperature interval of interest. With the FSS procedure~\cite{melchert2009autoscale}, we locate the critical temperature and extract the critical exponents for all these (quasi)order parameters as detailed in Table~\ref{Table: q=4, shell fine}. They agree with the existing numerical estimates of $\nu=0.722$ and $\Delta_\sigma=0.128$~\cite{salas1997logarithmic} very well. The deviation from the theoretical values of $\nu=2/3$ and $\Delta_\sigma=1/8$~\cite{wu1982potts} arises from the omission of the higher-order corrections. We conclude that all occupancies of these states along with the magnetization-like quantity $M_{\mathrm{map}}^{}$ can serve as alternative (quasi)order parameters for $q=4$ Potts model as well.

\section{Conclusion}
\label{sec: conclusion}
Through systematic finite-size scaling analyses, we have discovered and validated a family of alternative (quasi)order parameters for the square-lattice
$q=3$ and $q=4$ Potts models, grounded in the underlying principles of universality and symmetry. These include (1) the conventional order parameter based on dominant spin state occupancy, (2) order parameters derived from secondary and minimal spin states, and (3) magnetization-like quasiorder parameter emerging from reduced representations with only two spin components. We have rigorously established the intrinsic relationships between these alternative (quasi)order parameters, demonstrating that they collectively provide multiple consistent characterizations of critical behavior associated with different level of standard uncertainty. This unified framework explains the remarkable generalizability of Ising-trained machine-learning models to Potts systems. The models remain effective because criticality is redundantly encoded in these various (quasi)order parameters, persisting even when explicit spin state information is partially lost through representation reduction. 

Motivated by these findings, we further note that other quantities, such as $|N_1/N-1/q|$, may also serve as (quasi)order parameters. This corresponds to an even more drastic mapping in which all except one of the Potts states are mapped to zero. We find that any quantity, such as $M_\mathrm{map}$ and $M_1 \propto |N_1/N-1/q|$, that treats ordering in a subset of the Potts states differently from ordering in the other Potts states will have non-thermal fluctuations that increase as $T$ decreases and may obscure a peak in the thermal fluctuations near the critical point.

Our findings suggest that these (quasi)order parameters generalize to other $q$-state Potts models, different lattice geometries (e.g., triangular or 3D), and systems with extended interactions (anti-ferromagnetic couplings, next-nearest neighbors). They may further enable studies of real-time dynamics or non-equilibrium phase transitions. More broadly, this work provides a concrete example of how reduced spin representations can preserve the essential thermodynamic information required for identifying criticality, highlighting that conventional order parameters are not unique descriptors of critical phenomena and pointing toward the possibility that analogous quasiorder parameters, and the representations implicitly used by machine-learning models, may arise in a wider class of interacting spin systems.

% Crucially, this work bridges machine learning's empirical success in phase classification with fundamental thermodynamic quantities, demonstrating that conventional order parameters are not unique descriptors of critical phenomena. Future work could explore these generalizations systematically or analyze how neural networks leverage such (quasi)order parameters across diverse spin systems.

%%%%%%%%%%%%%%%%%%%%%%%%%%%%%%%%%%%%%%%%%%%%%%%%%%
\section*{Acknowledgments}
% We acknowledge for \dots stimulating discussions. 
Y.D. thanks Ke Xu for the assistance with the usage of HPC in the cluster. This work was supported by the University of Bergen through ``Akademiaavtalen'' (K.T. and N.S.), the Taishan Scholars Program under Grant No. TSQNZ20221162, and the Shandong Excellent Young Scientists Fund Program (Overseas) under Grant No. 2023HWYQ-106 (Y.D.). Y.D. thanks the support from the cluster in FIAS and the Norwegian e-infrastructure UNINETT Sigma2 with Projects No. NS9753K and No. NN9753K for the HPC resources and data storage in Norway. The work of N.S. was done at FIAS, supported by the AI grant at FIAS of SAMSON AG.

% The computations/simulations/[SIMILAR] were performed on resources provided by
% UNINETT Sigma2 - the National Infrastructure for High Performance Computing and
% Data Storage in Norway
%%%%%%%%%%%%%%%%%%%%%%%%%%%%%%%%%%%%%%%%%%%%%%%%%%

%\clearpage

\appendix

\renewcommand{\thefigure}{\thesection\arabic{figure}} 
\setcounter{figure}{0}

%\onecolumn

% \twocolumn[
\section{Finite-size scaling of alternative (quasi)order parameters for $q=3$ and $q=4$ Potts model on a square lattice}
% \onecolumn
% ]
% \subsection{$q=3$ Potts model on a square lattice}
\label{sec: appendix}

\begin{widetext}

In this Appendix, we present additional finite-size scaling analyses for the alternative (quasi)order parameters introduced in the main text for the $q=3$ and $q=4$ Potts models on a square lattice. These results complement the main discussion by providing detailed numerical evidence for the scaling behavior with known theoretical or numerical critical temperatures and critical exponents, thereby supporting the robustness of our conclusions.

\begin{figure*}[!htb] 
\centering 
\begin{minipage}{0.33\linewidth}
\centering 
\includegraphics[width=2.3in]{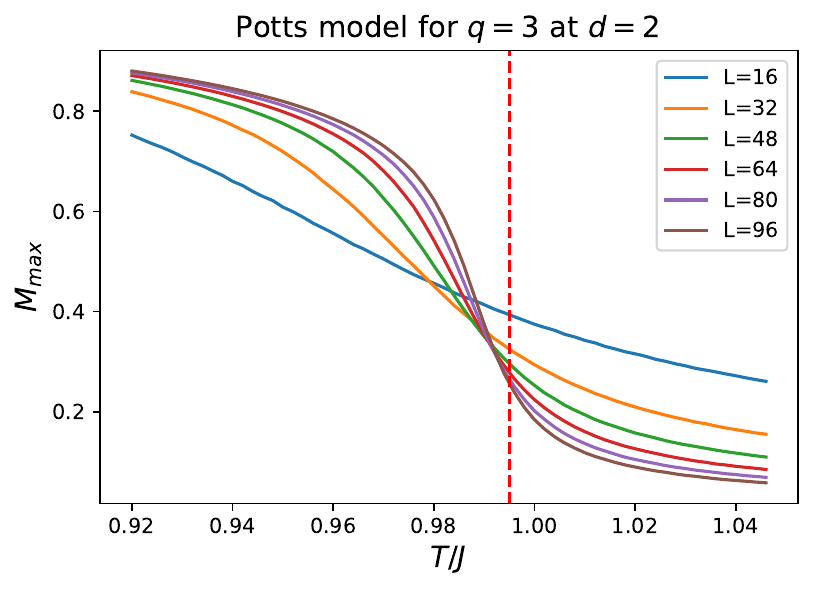}  
\end{minipage}
\begin{minipage}{0.33\linewidth}
\centering
\includegraphics[width=2.3in]{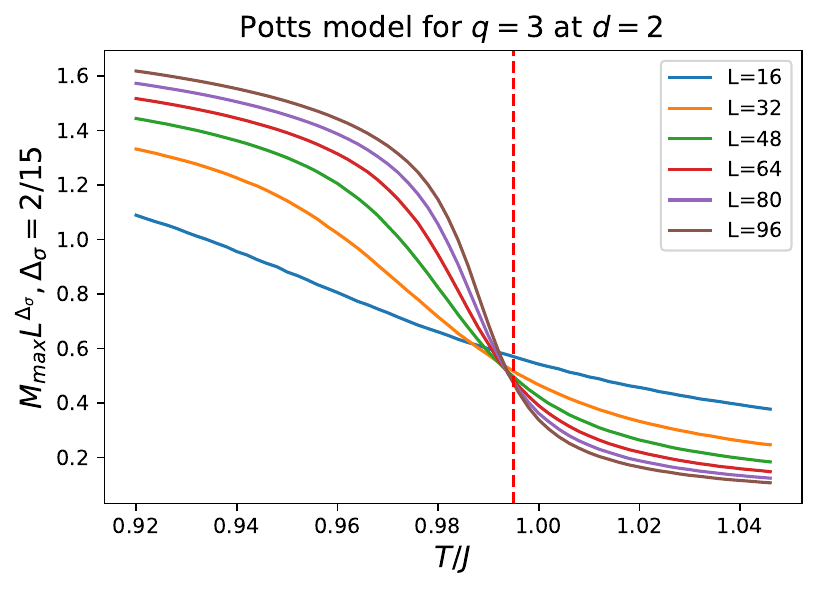} 
\end{minipage}
\begin{minipage}{0.32\linewidth}
\centering 
\includegraphics[width=2.3in]{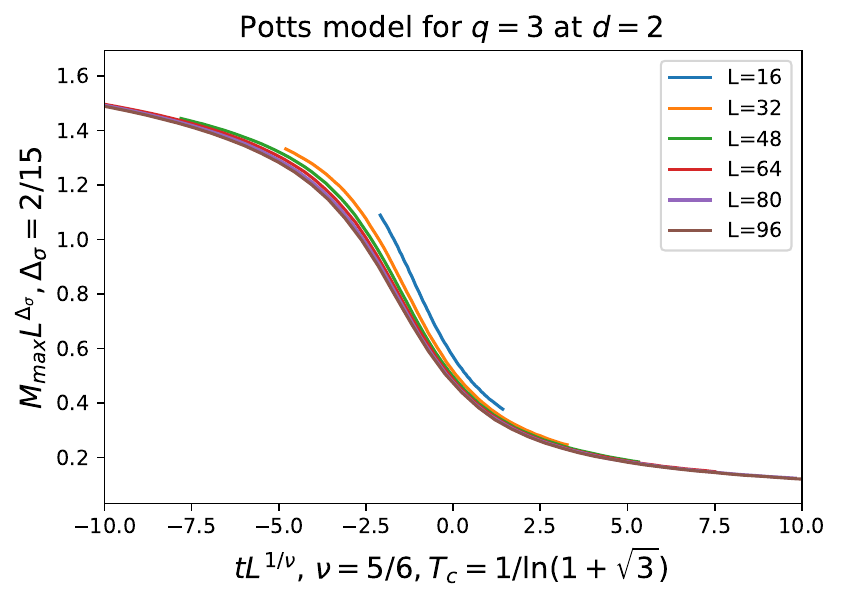} 
\end{minipage}
\caption{(Color online) (Left) Order parameter $M_{\mathrm{max}}$ as a function of temperature $T$ for the $q=3$ Potts model on a square lattice with six different system sizes $L$. (Middle) Scaled order parameter $M_{\mathrm{max}}L^{\Delta_\sigma}$ versus $T$, with curves intersecting near the theoretical critical temperature $T_c$, marked by the red vertical dashed line. (Right) Scaled order parameter $M_{\mathrm{max}}L^{\Delta_\sigma}$ against the rescaled temperature $tL^{1/\nu}$, demonstrating data collapse consistent with finite-size scaling.}
\label{Fig: Mmax}
\end{figure*}

\begin{figure*}[!htb] 
\centering 
\begin{minipage}{0.33\linewidth} 
\centering 
\includegraphics[width=2.3in]{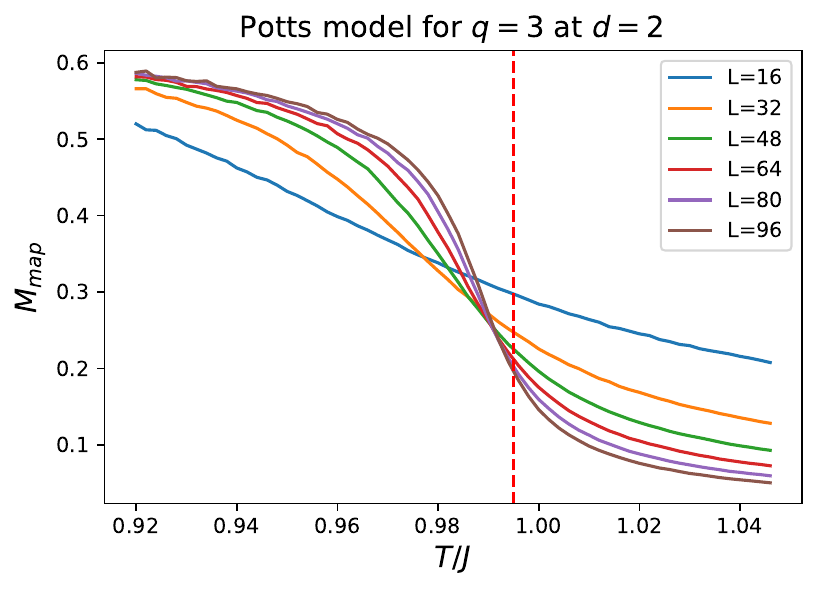}  
\end{minipage}%
\begin{minipage}{0.33\linewidth} 
\centering 
\includegraphics[width=2.3in]{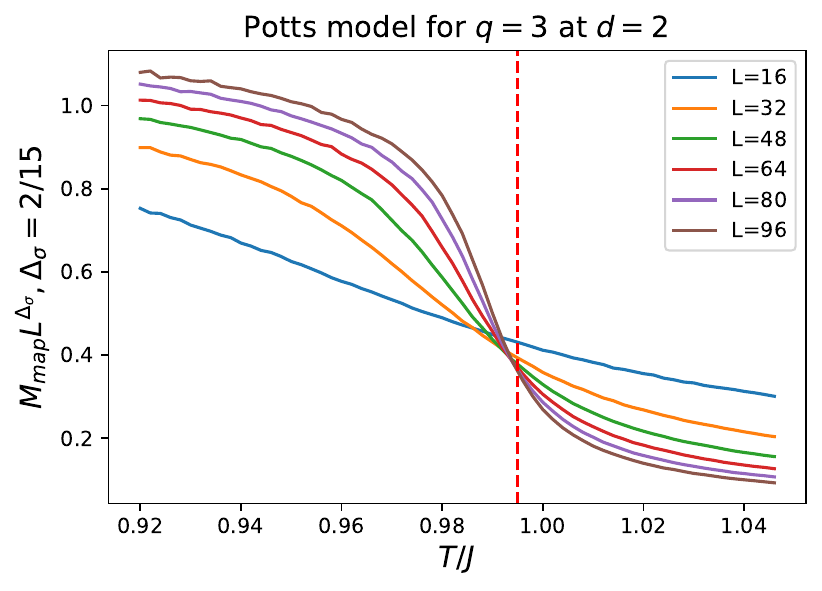} 
\end{minipage} 
\begin{minipage}{0.33\linewidth} 
\centering 
\includegraphics[width=2.3in]{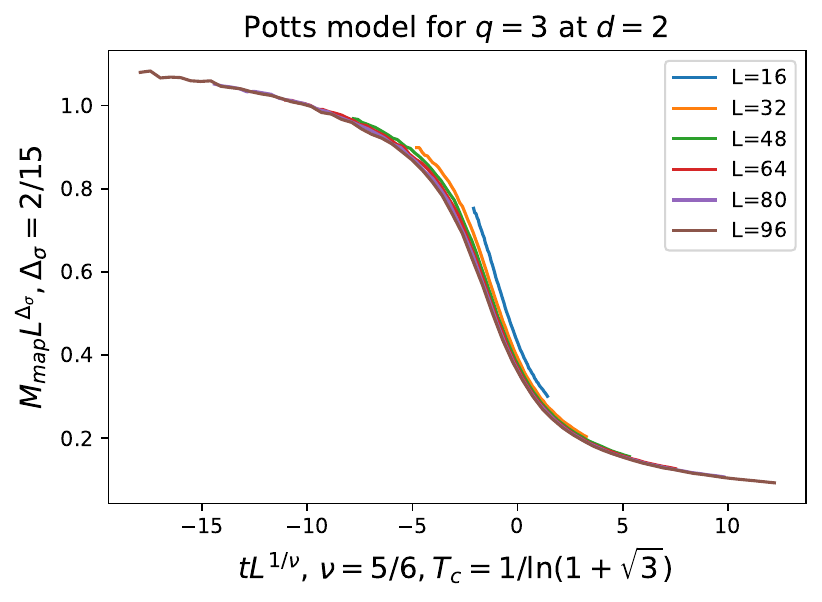} 
\end{minipage} 
\caption{(Color online) (Left) Quasiorder parameter $M_{\mathrm{map}}$ as a function of temperature $T$ for the $q=3$ Potts model on a square lattice with six different system sizes $L$. (Middle) Scaled quasiorder parameter $M_{\mathrm{map}}L^{\Delta_\sigma}$ versus $T$, with curves intersecting near the theoretical critical temperature $T_c$, marked by the red vertical dashed line. (Right) Scaled quasiorder parameter $M_{\mathrm{map}}L^{\Delta_\sigma}$ against the rescaled temperature $tL^{1/\nu}$, demonstrating data collapse consistent with finite-size scaling.}
\label{Fig: Mmap}
\end{figure*}

\begin{figure*}[!htb] 
\centering 
\begin{minipage}{0.33\linewidth} 
\centering 
\includegraphics[width=2.3in]{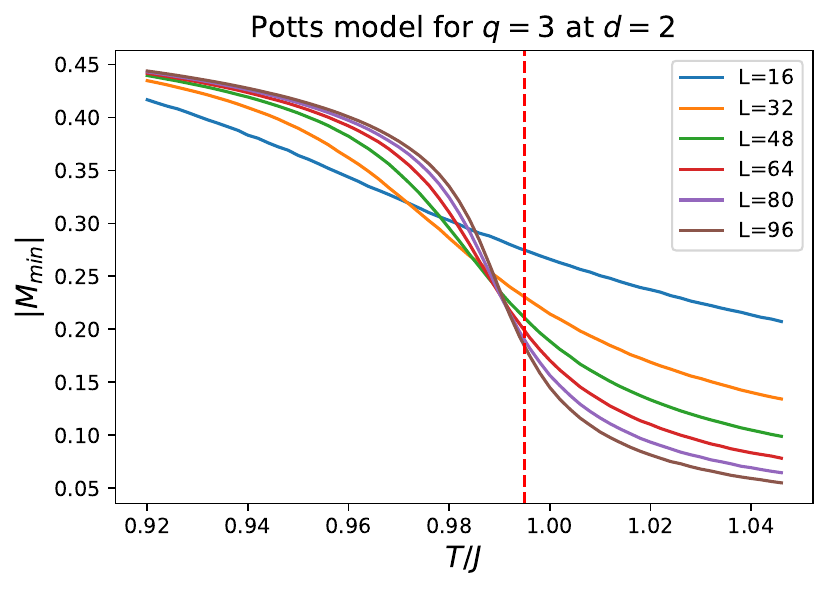} 
\end{minipage}%
\begin{minipage}{0.33\linewidth} 
\centering 
\includegraphics[width=2.3in]{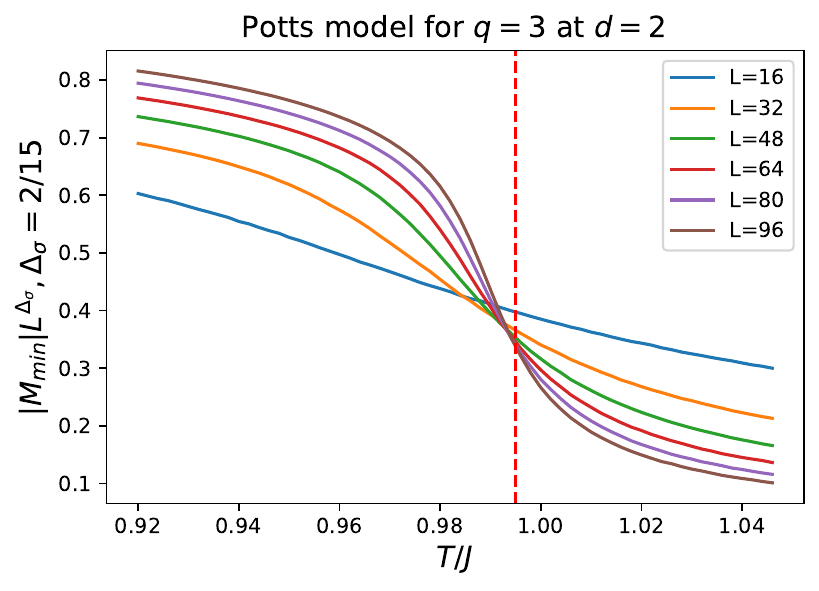} 
\end{minipage} 
\begin{minipage}{0.33\linewidth} 
\centering 
\includegraphics[width=2.3in]{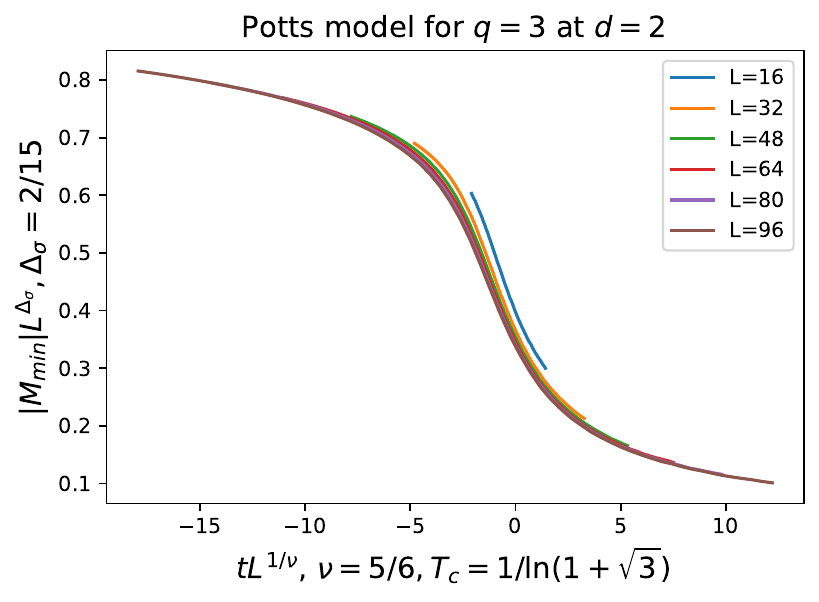} 
\end{minipage} 
\caption{(Color online) (Left) Order parameter $M_{\mathrm{min}}$ as a function of temperature $T$ for the $q=3$ Potts model on a square lattice with six different system sizes $L$. (Middle) Scaled order parameter $M_{\mathrm{min}}L^{\Delta_\sigma}$ versus $T$, with curves intersecting near the theoretical critical temperature $T_c$, marked by the red vertical dashed line. (Right) Scaled order parameter $M_{\mathrm{min}}L^{\Delta_\sigma}$ against the rescaled temperature $tL^{1/\nu}$, demonstrating data collapse consistent with finite-size scaling.}
\label{Fig: Mmin}
\end{figure*}

\begin{figure*}[h] 
\centering 
\begin{minipage}{0.33\linewidth} 
\centering 
\includegraphics[width=2.3in]{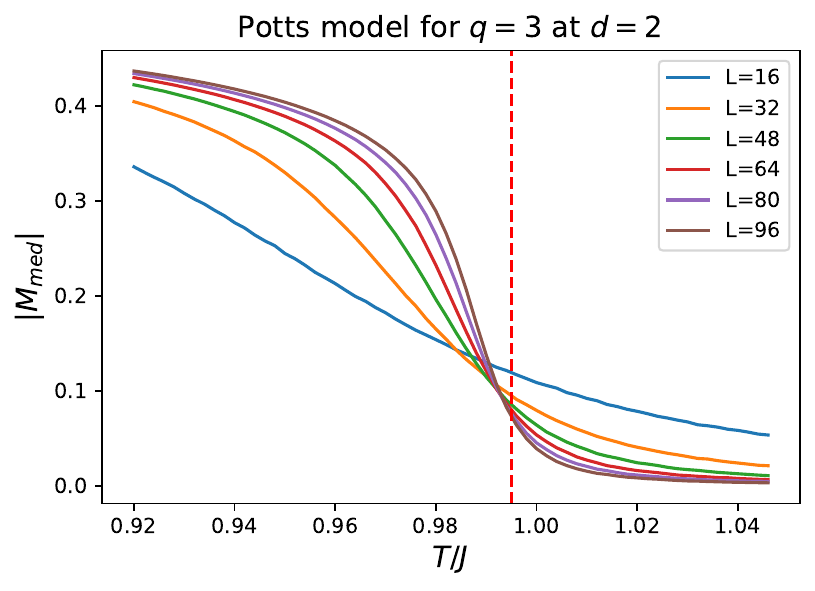} 
\end{minipage}%
\begin{minipage}{0.33\linewidth} 
\centering 
\includegraphics[width=2.3in]{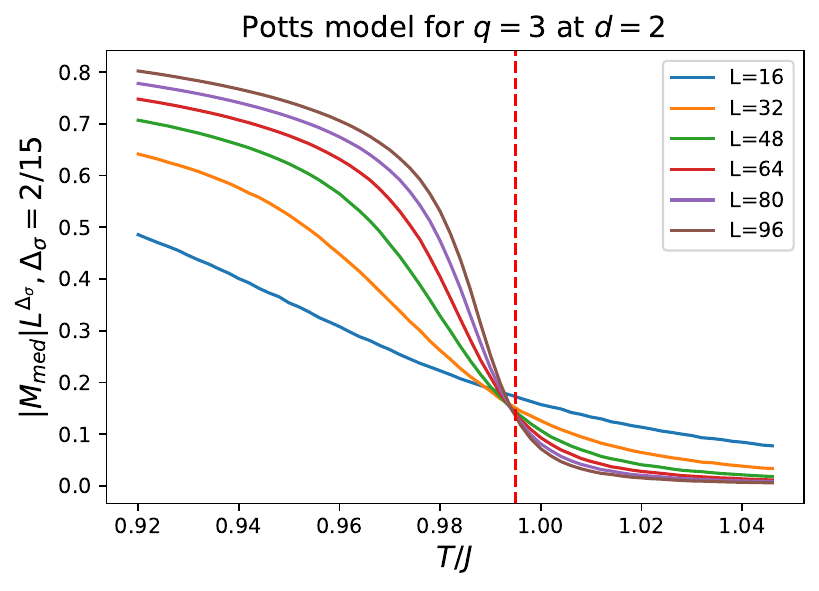} 
\end{minipage} 
\begin{minipage}{0.33\linewidth} 
\centering 
\includegraphics[width=2.3in]{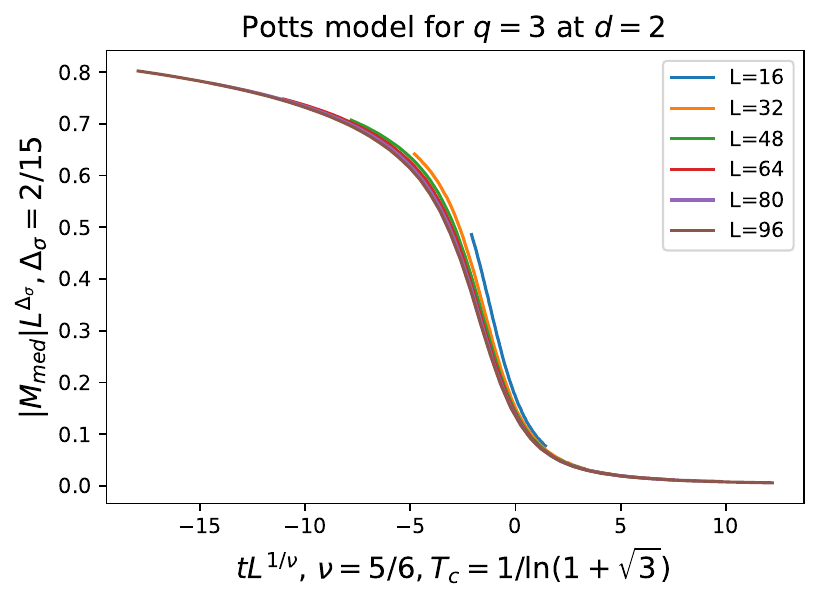} 
\end{minipage} 
\caption{(Color online) (Left) Order parameter $M_{\mathrm{med}}$ as a function of temperature $T$ for the $q=3$ Potts model on a square lattice with six different system sizes $L$. (Middle) Scaled order parameter $M_{\mathrm{med}}L^{\Delta_\sigma}$ versus $T$, with curves intersecting near the theoretical critical temperature $T_c$, marked by the red vertical dashed line. (Right) Scaled order parameter $M_{\mathrm{med}}L^{\Delta_\sigma}$ against the rescaled temperature $tL^{1/\nu}$, demonstrating data collapse consistent with finite-size scaling.}
\label{Fig: Mmed}
\end{figure*}

%\clearpage
% \subsection{$q=4$ Potts model on a square lattice}

\begin{figure*}[h] 
\centering 
\begin{minipage}{0.33\linewidth} 
\centering 
\includegraphics[width=2.3in]{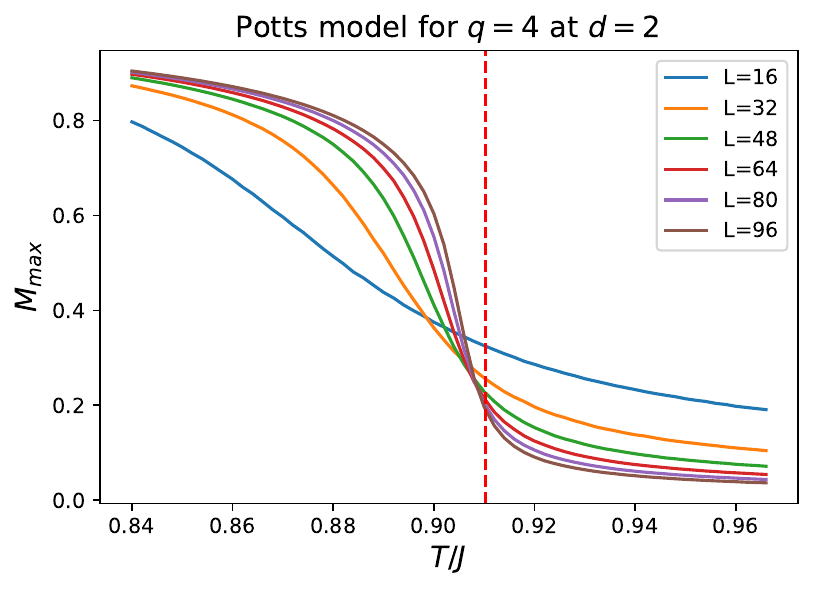}  
\end{minipage}%
\begin{minipage}{0.33\linewidth} 
\centering 
\includegraphics[width=2.3in]{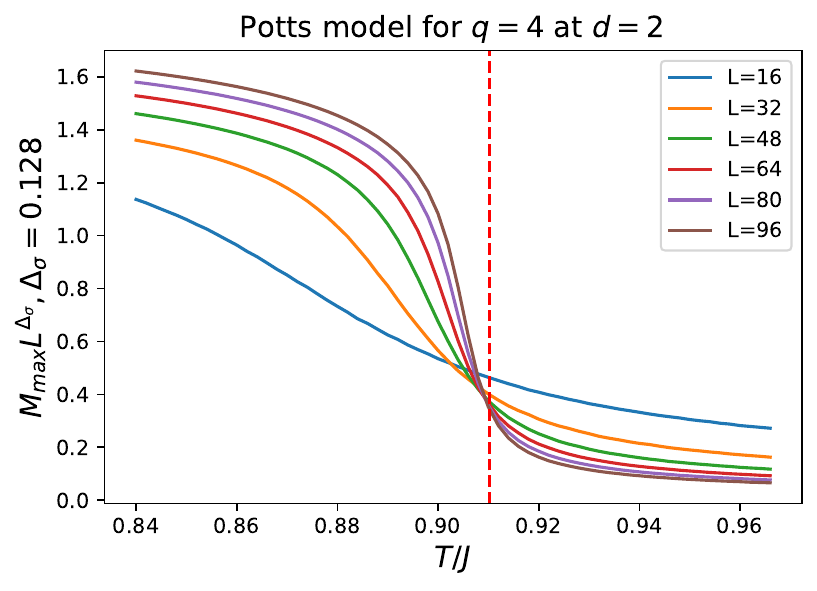} 
\end{minipage} 
\begin{minipage}{0.33\linewidth} 
\centering 
\includegraphics[width=2.3in]{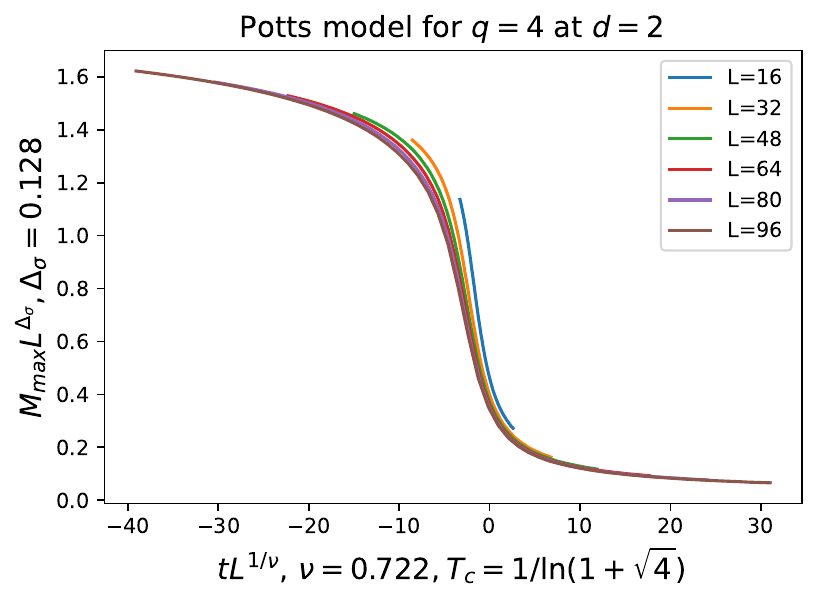} 
\end{minipage} 
\caption{(Color online) (Left) Order parameter $M_{\mathrm{max}}$ as a function of temperature $T$ for the $q=4$ Potts model on a square lattice with six different system sizes $L$. (Middle) Scaled order parameter $M_{\mathrm{max}}L^{\Delta_\sigma}$ versus $T$, with curves intersecting near the theoretical critical temperature $T_c$, marked by the red vertical dashed line. (Right) Scaled order parameter $M_{\mathrm{max}}L^{\Delta_\sigma}$ against the rescaled temperature $tL^{1/\nu}$, demonstrating data collapse consistent with finite-size scaling.}
\label{Fig: Mmaxq4}
\end{figure*}

\begin{figure*}[!htb] 
\centering 
\begin{minipage}{0.33\linewidth} 
\centering 
\includegraphics[width=2.3in]{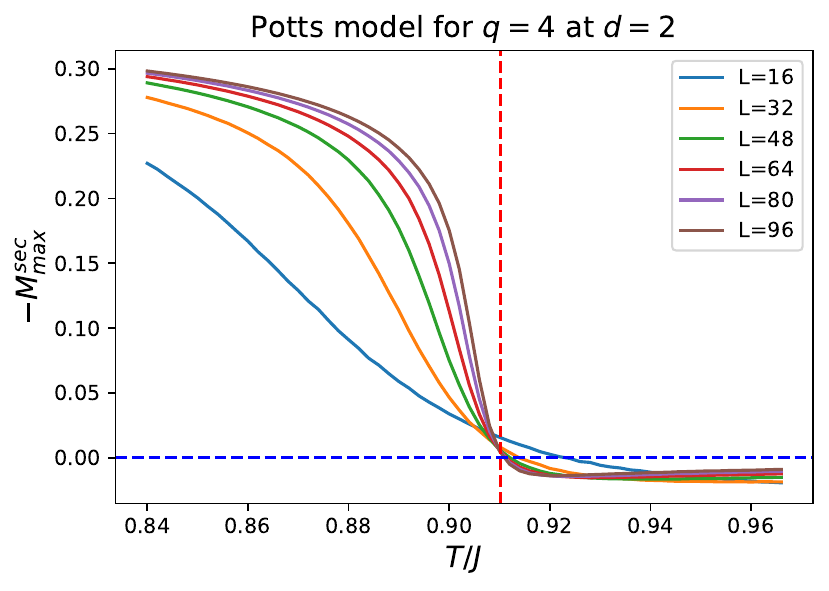}  
\end{minipage}%
\begin{minipage}{0.33\linewidth} 
\centering 
\includegraphics[width=2.3in]{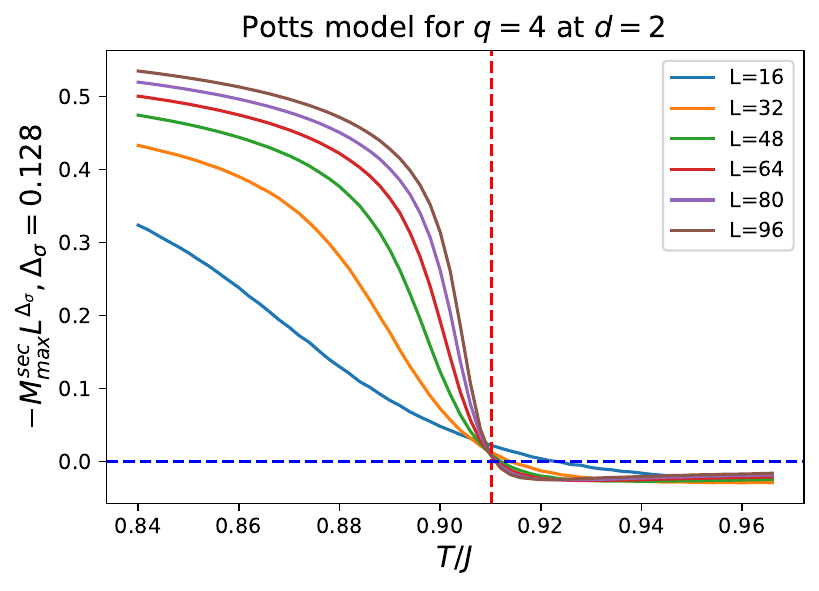} 
\end{minipage} 
\begin{minipage}{0.33\linewidth} 
\centering 
\includegraphics[width=2.3in]{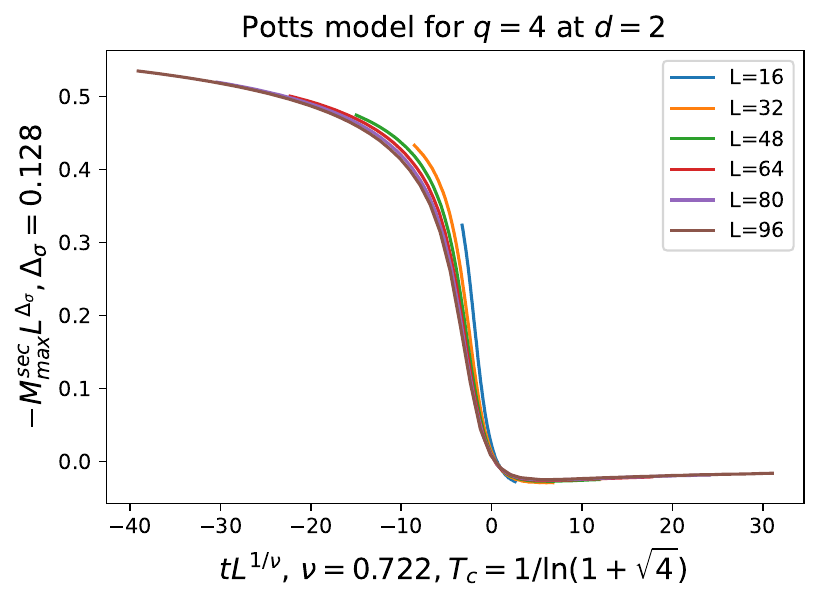} 
\end{minipage} 
\caption{(Color online) (Left) Order parameter $M^{\mathrm{sec}}_{\mathrm{max}}$ as a function of temperature $T$ for the $q=4$ Potts model on a square lattice with six different system sizes $L$. The blue horizontal dashed line marks zero to highlight the sign change. (Middle) Scaled order parameter $M^{\mathrm{sec}}_{\mathrm{max}}L^{\Delta_\sigma}$ versus $T$, with curves intersecting near the theoretical critical temperature $T_c$, marked by the red vertical dashed line. (Right) Scaled order parameter $M^{\mathrm{sec}}_{\mathrm{max}}L^{\Delta_\sigma}$ against the rescaled temperature $tL^{1/\nu}$, demonstrating data collapse consistent with finite-size scaling.}
\label{Fig: Msecmaxq4}
\end{figure*}

\begin{figure*}[!htb] 
\centering 
\begin{minipage}{0.33\linewidth} 
\centering 
\includegraphics[width=2.3in]{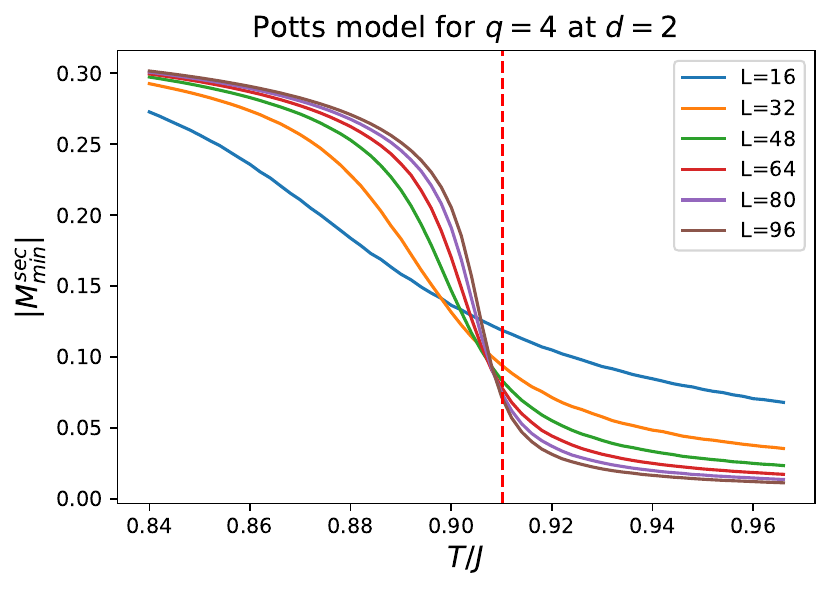} 
\end{minipage}%
\begin{minipage}{0.33\linewidth} 
\centering 
\includegraphics[width=2.3in]{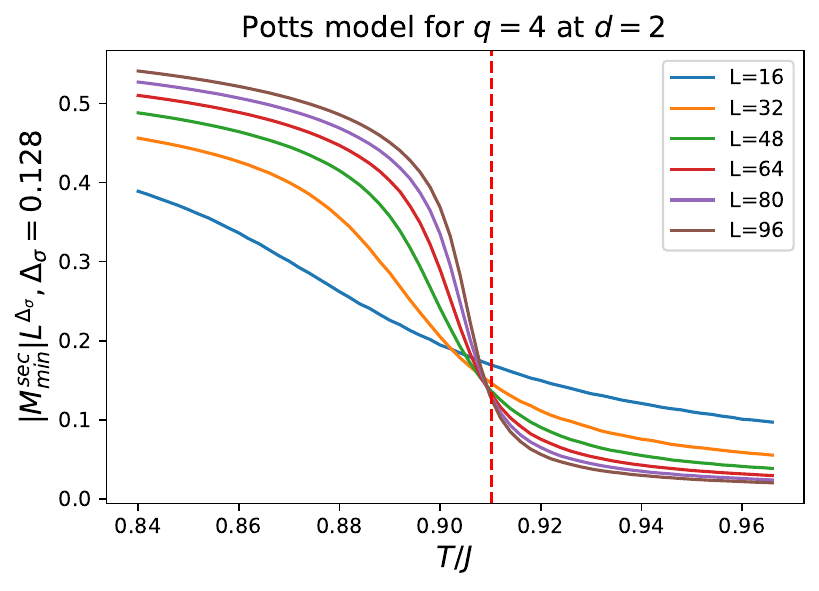} 
\end{minipage} 
\begin{minipage}{0.33\linewidth} 
\centering 
\includegraphics[width=2.3in]{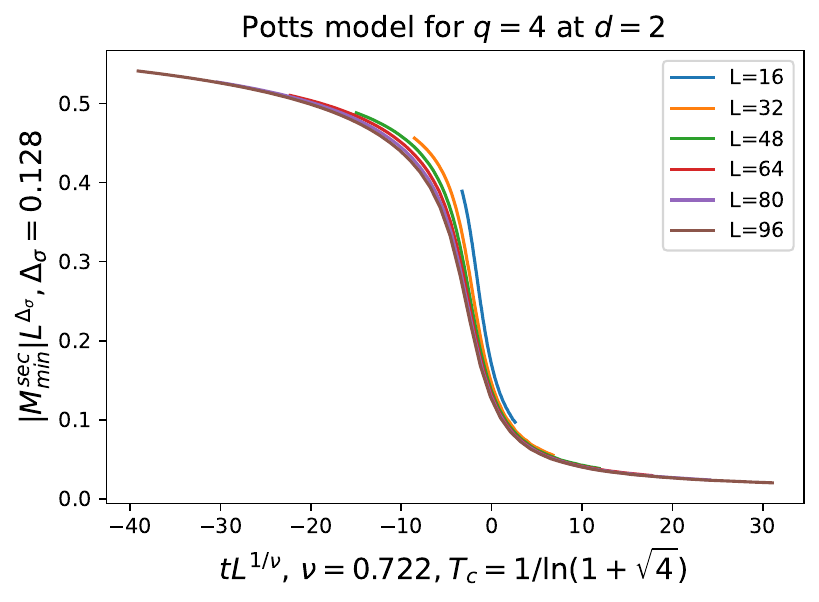} 
\end{minipage} 
\caption{(Color online) (Left) Order parameter $M^{\mathrm{sec}}_{\mathrm{min}}$ as a function of temperature $T$ for the $q=4$ Potts model on a square lattice with six different system sizes $L$. (Middle) Scaled order parameter $M^{\mathrm{sec}}_{\mathrm{min}}L^{\Delta_\sigma}$ versus $T$, with curves intersecting near the theoretical critical temperature $T_c$, marked by the red vertical dashed line. (Right) Scaled order parameter $M^{\mathrm{sec}}_{\mathrm{min}}L^{\Delta_\sigma}$ against the rescaled temperature $tL^{1/\nu}$, demonstrating data collapse consistent with finite-size scaling.}
\label{Fig: Msecminq4}
\end{figure*}

\begin{figure*}[!htb] 
\centering 
\begin{minipage}{0.33\linewidth} 
\centering 
\includegraphics[width=2.3in]{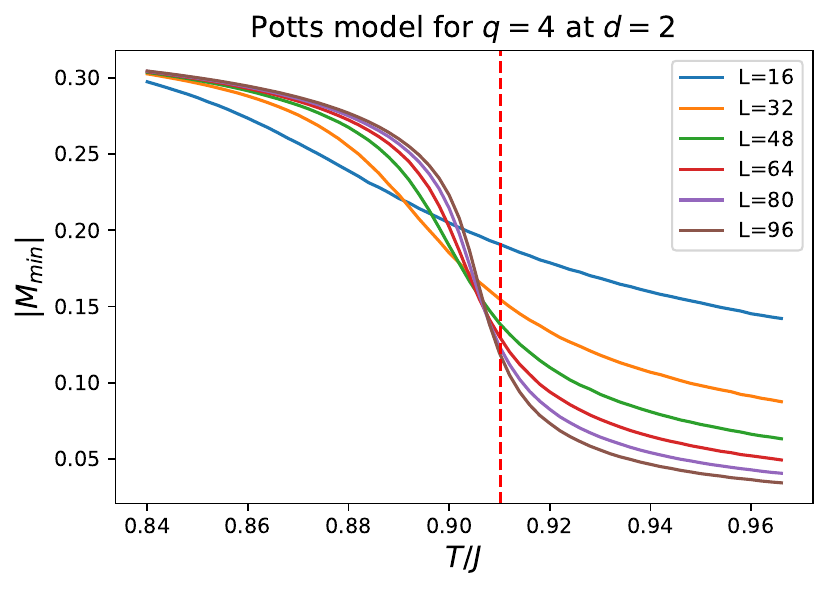} 
\end{minipage}%
\begin{minipage}{0.33\linewidth} 
\centering 
\includegraphics[width=2.3in]{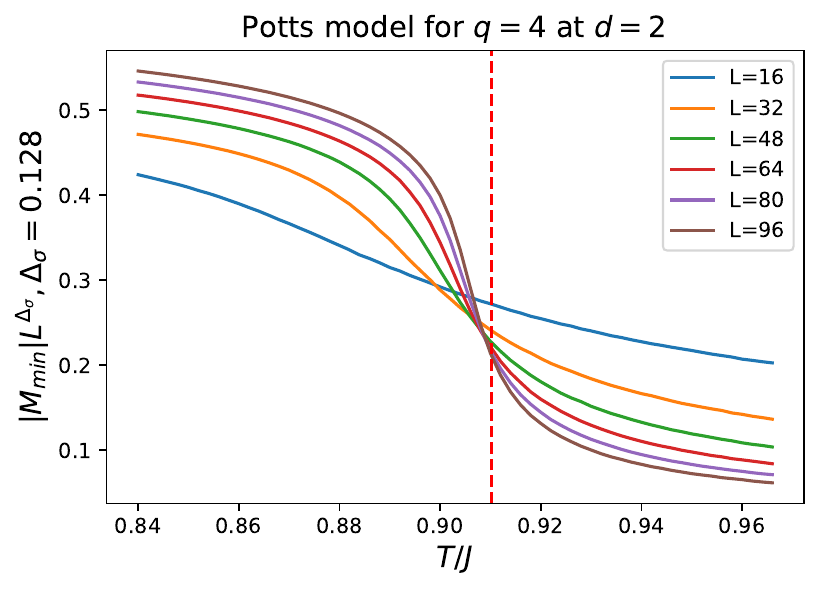} 
\end{minipage} 
\begin{minipage}{0.33\linewidth} 
\centering 
\includegraphics[width=2.3in]{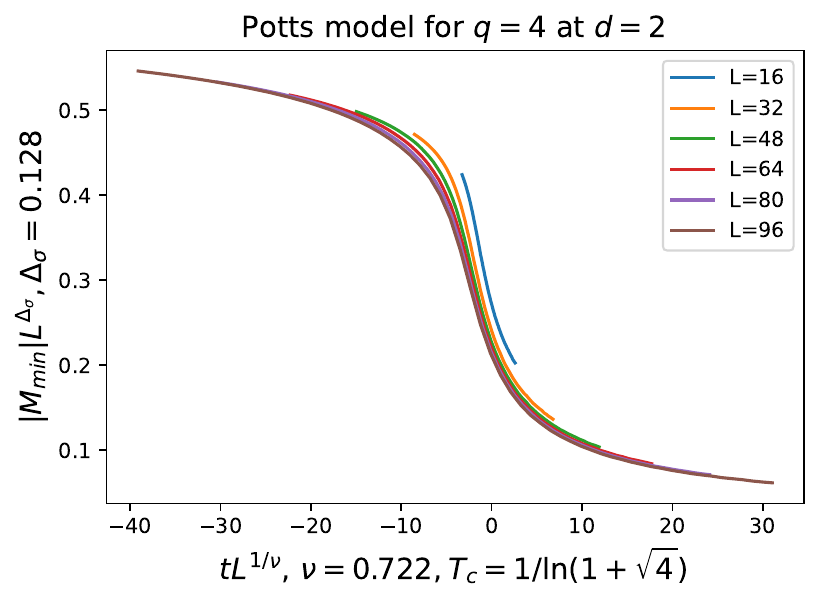} 
\end{minipage} 
\caption{(Color online) (Left) Order parameter $M_{\mathrm{min}}$ as a function of temperature $T$ for the $q=4$ Potts model on a square lattice with six different system sizes $L$. (Middle) Scaled order parameter $M_{\mathrm{min}}L^{\Delta_\sigma}$ versus $T$, with curves intersecting near the theoretical critical temperature $T_c$, marked by the red vertical dashed line. (Right) Scaled order parameter $M_{\mathrm{min}}L^{\Delta_\sigma}$ against the rescaled temperature $tL^{1/\nu}$, demonstrating data collapse consistent with finite-size scaling.}
\label{Fig: Mminq4}
\end{figure*}

\begin{figure*}[!htb] 
\centering 
\begin{minipage}{0.32\linewidth} 
\centering 
\includegraphics[width=2.3in]{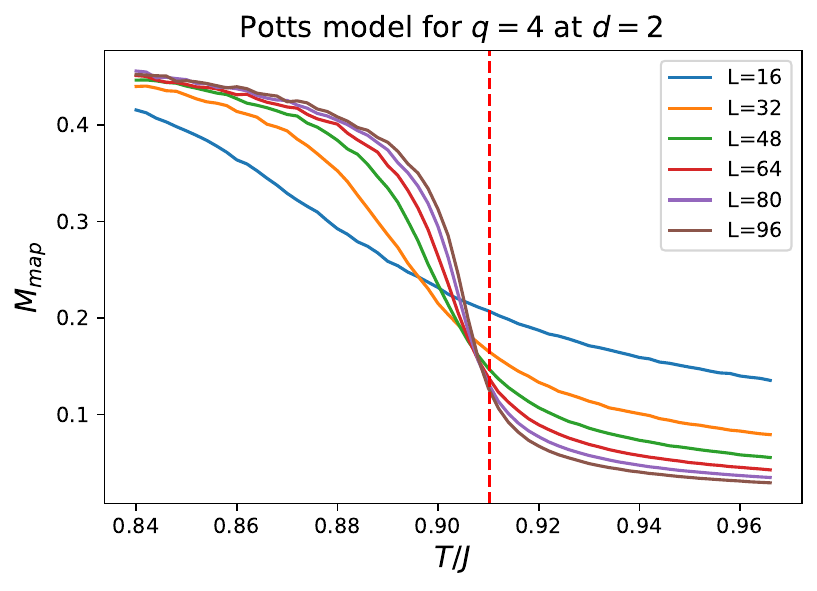} 
\end{minipage}%
\begin{minipage}{0.32\linewidth} 
\centering 
\includegraphics[width=2.3in]{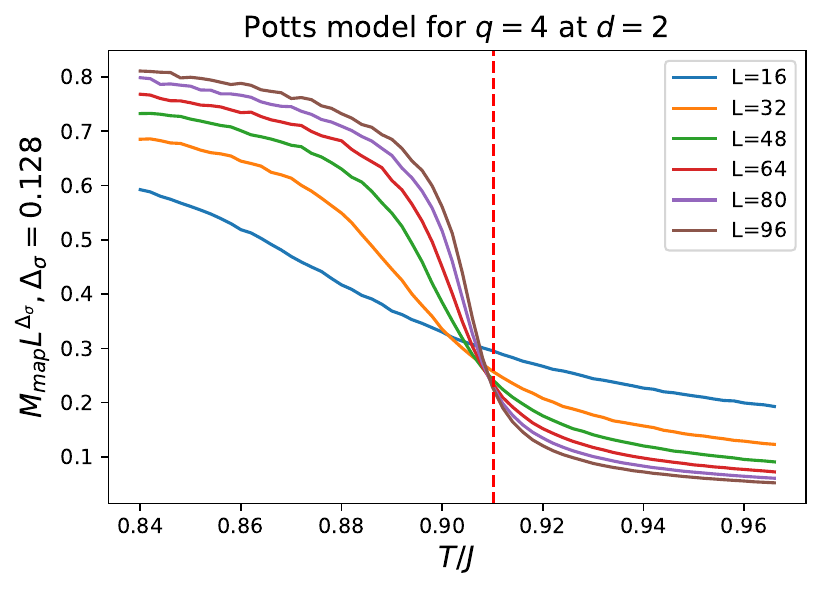} 
\end{minipage} 
\begin{minipage}{0.32\linewidth} 
\centering 
\includegraphics[width=2.3in]{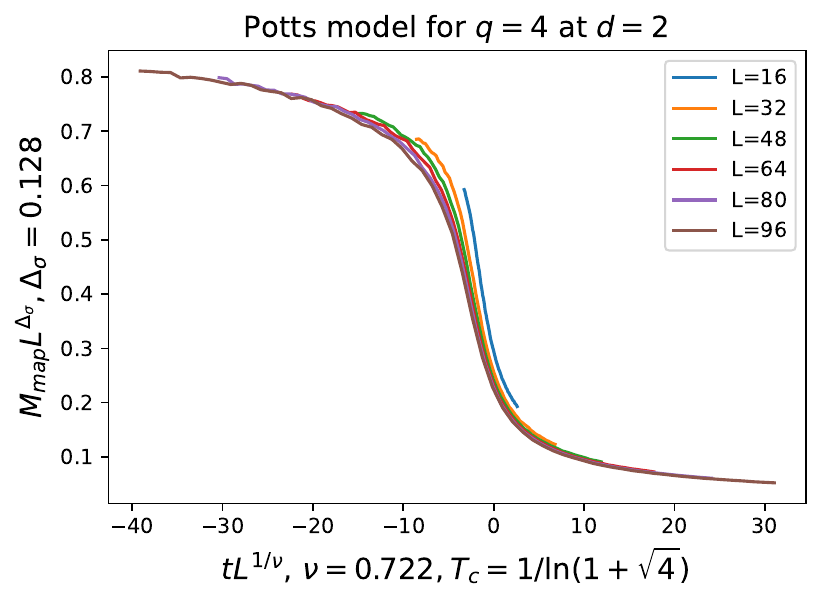} 
\end{minipage} 
\caption{(Color online) (Left) Quasiorder parameter $M_{\mathrm{map}}$ as a function of temperature $T$ for the $q=4$ Potts model on a square lattice with six different system sizes $L$. (Middle) Scaled quasiorder parameter $M_{\mathrm{map}}L^{\Delta_\sigma}$ versus $T$, with curves intersecting near the theoretical critical temperature $T_c$, marked by the red vertical dashed line. (Right) Scaled quasiorder parameter $M_{\mathrm{map}}L^{\Delta_\sigma}$ against the rescaled temperature $tL^{1/\nu}$, demonstrating data collapse consistent with finite-size scaling.}
\label{Fig: Mmapq4}
\end{figure*}
\end{widetext}

% %\bibliography{mtjet.bib}
% \begin{thebibliography}{99}
  
% \end{thebibliography}

\clearpage

%\section*{Bibliography}
%\bibliographystyle{apsrev4-2}
% \bibliographystyle{JHEP}
\bibliography{duyl}

\end{document}